\documentclass[%
 reprint,
 amsmath,amssymb,
 aps,
prb,
]{revtex4-1}
\usepackage{algorithm}
\usepackage{algpseudocode}
\usepackage{graphicx}
\usepackage{dcolumn}
\usepackage{bm}
\usepackage{multirow}
\usepackage{color}
\usepackage{textcomp}

\def\br{{\bm r}}

\def\vii{{\bm v}}

\def\ace{\varphi}

\def\calF{\mathcal{F}}

\def\rc{r_{\text{cut}}}

\def\lACE{{ACE$^{\text{(l)}}$\,}}
\def\gACE{{ACE$^{\text{(g)}}$\,}}

\newcommand {\braket}[2]{ \langle #1 | #2  \rangle }
\newcommand {\braHket}[3]{ \langle #1 | #2 | #3 \rangle }
\newcommand {\ket}[1]{| #1 \rangle}
\newcommand {\bra}[1]{\langle  #1 | }

\newcommand {\pder}[2]{\frac{\partial  #1}{\partial #2} }

\DeclareMathOperator*{\SumInt}{%
\mathchoice%
  {\ooalign{$\displaystyle\sum$\cr\hidewidth$\displaystyle\int$\hidewidth\cr}}
  {\ooalign{\raisebox{.14\height}{\scalebox{.7}{$\textstyle\sum$}}\cr\hidewidth$\textstyle\int$\hidewidth\cr}}
  {\ooalign{\raisebox{.2\height}{\scalebox{.6}{$\scriptstyle\sum$}}\cr$\scriptstyle\int$\cr}}
  {\ooalign{\raisebox{.2\height}{\scalebox{.6}{$\scriptstyle\sum$}}\cr$\scriptstyle\int$\cr}}
}
\begin{document}

\title{Graph Atomic Cluster Expansion for semilocal interactions beyond equivariant message passing}

\author{Anton Bochkarev}
\affiliation{%
 ICAMS, Ruhr-Universit\"at Bochum, Bochum, Germany
}%

\author{Yury Lysogorskiy}
\affiliation{%
 ICAMS, Ruhr-Universit\"at Bochum, Bochum, Germany
}%

\author{Ralf Drautz}
\affiliation{%
 ICAMS, Ruhr-Universit\"at Bochum, Bochum, Germany
}%

\date{\today}

\begin{abstract}
The Atomic Cluster Expansion provides local, complete basis functions that enable efficient parametrization of many-atom interactions. We extend the Atomic Cluster Expansion to incorporate graph basis functions. This naturally leads to representations that enable the efficient description of semilocal interactions in physically and chemically transparent form. Simplification of the graph expansion by tensor decomposition results in an iterative procedure that comprises current message-passing machine learning interatomic potentials.
We demonstrate the accuracy and efficiency of the graph Atomic Cluster Expansion for a number of small molecules, clusters and a general-purpose model for carbon. We further show that the graph Atomic Cluster Expansion scales linearly with number of neighbors and layer depth of the graph basis functions.

\end{abstract}

\maketitle

\section{Introduction}

Semilocal interatomic interactions beyond the reach of Hamiltonian matrix elements $\braHket{i\alpha}{\hat{H}}{j \beta}$ between orbitals $\alpha$, $\beta$ located on atoms $i$ and $j$ are ubiquitous in quantum mechanical calculations. Semilocal interactions involve contributions of different origin. First, diagonalization of the Hamiltonian induces interactions that reach multiple times beyond that of the Hamiltonian matrix, as can be seen from a simple series expansion that results in contributions of the form $ \sum_{k \gamma} \braHket{i\alpha}{\hat{H}}{k \gamma}\braHket{k\gamma}{\hat{H}}{j \beta}$ and higher order \cite{Drautz06}. Second, relaxation of the electronic structure in self-consistent calculations induces interactions beyond the direct range of the Hamiltonian matrix elements \cite{Bochkarev2022mlACE}. Third, direct interactions induced by electronic correlations, dispersion corrections in density functional theory calculations, extend beyond the reach of Hamiltonian matrix elements.

We base our analysis of local and semilocal contributions on the Atomic Cluster Expansion (ACE) that enables accurate and efficient parametrization of many-atom interactions. The basis functions of ACE are complete and can represent other local descriptors\cite{Drautz19,Dusson2022}. For example, the smooth overlap of atomic positions (SOAP) descriptor \cite{Csanyi2013SOAP}, the Spectral Neighbor Analysis Potential (SNAP) \cite{Thompson2015SNAP}, the atom-centred symmetry functions (ACSF) \cite{Behler2007ACSF} and many other descriptors can be cast in the form of ACE. By expanding Cartesian in spherical coordinates, other models such as the moment tensor potentials (MTP)\cite{Shapeev2016MTP} can also be represented \cite{Drautz2020,Dusson2022}.

ACE is efficient to evaluate and linear-scaling with the number of basis functions irrespective of the body order of the expansion\cite{Lysogorskiy2021PACE,Kaliuzhnyi2022recursive}. This means that higher body-order interactions are captured efficiently and machine learning frameworks for non-linearly transforming descriptors to energies, such as neural network potentials, Gaussian process regression or kernel methods, are no longer necessary but optional for achieving accurate models. In fact, it was demonstrated in a number of publications that ACE exceeds the accuracy and numerical efficiency of more traditional machine learning interatomic potentials \cite{Qamar2023,Kovacs2021linACE,Lysogorskiy2021PACE,Ibrahim23}. 

The ACE formalism allows for the seamless incorporation of additional features, such as atomic magnetic moments and atomic or orbital charges, as well as the representation of vectorial and tensorial outputs \cite{Drautz2020,Rinaldi2023noncollinear}. ACE is neither bound to atomic interactions nor three-dimensional space and was extended to jet tagging \cite{Munoz2022}, Hamiltonian matrix representations \cite{Zhang2022equivariantACE} and wave functions \cite{Drautz2022ACEwave,Zhou2023multilevel}. The fact that ACE builds on a basis representation enables efficient uncertainty prediction and active learning and exploration \cite{Lysogorskiy2023active,Oor2023hyper}. ACE may therefore been seen to provide a general and unified representation of atom-centered approaches \cite{Behler2007ACSF,Csanyi2010GAP, Shapeev2016MTP,Braams2009PIPs,Faber2018FCHL,Zhang2018DeepPot,Smith2017ANINN,Zaverkin2020}. 

In the past years graph and message-passing representations \cite{Klicpera2020DimeNet,Anderson2019Cormorant,Lubbers2018,Thomas2018TensorField,Batzner2022nequip,Satorras2021EGNN,Unke2019Physnet,Schutt2017schnet,Haghighatlari2021newtonnet,Schutt2021Painn,Klicpera2022gemnet,Chmiela2017machine,Musaelian2022Allegro,Pozdnyakov2023smooth} were developed in parallel to atom-centered representations and only recently unified views on atom-centred and message-passing approaches emerged\cite{Nigam2022unified,Batatia2022design,Bochkarev2022mlACE}. Here ACE-based message passing \cite{batatia2023general} provided the most accurate models \cite{batatia2023mace,kovacs2023evaluation} as these incorporated semilocal interactions \cite{Bochkarev2022mlACE,Drautz06,thomas2021rigorous}. On each atom messages in the form of ACE are evaluated and passed to neighboring atoms, which use these to construct the next layer of ACE messages until after some layers energy and forces are evaluated. Each message thereby condenses the representation of atomic environments and one has to ensure that the iteration of messages is as descriptive as possible.

Here we extend local ACE to incorporate graphs. We show that graph ACE encompasses atom-centred local ACE as well as multi-atom and multi-layer message-passing architectures. We demonstrate that ACE generalizes and outperforms current local and semilocal machine learning interatomic potentials with respect to accuracy and efficiency.
We start by introducing graph-based cluster basis functions in Sec.~\ref{sec:gACE}. We then show that invariance and equivariance with respect to translation, rotation, inversion and permutation can be achieved along the same lines of local ACE in Sec.~\ref{sec:equi} before we compare local and global, graph ACE in detail in Sec.~\ref{sec:localglobal}. Here we use the term global as the graph basis functions can in principle extend over all atoms, but as we will see this does not imply that the number of atoms in a model is fixed. We discuss the quantum mechanical foundation of graph ACE in Sec.~\ref{sec:qmfoundation}. The global, graph ACE is significantly more complex than local ACE and we introduce a series of simplifications in Sec.~\ref{sec:simple}. This leads to recursive evaluation within the dandelion approximation in Sec.~\ref{sec:dandelion}. We then show that the dandelion approximation is closely related to message-passing architectures with multi-atom messages and in fact provides a derivation of the multi-ACE framework in Sec.~\ref{sec:comparison}. After having established that message-passing architectures can be viewed as a recursive evaluation of graph ACE, we further show that graph ACE exceeds the numerical accuracy and efficiency of other potentials for molecules, clusters and solids in Sec.~\ref{sec:application}. We conclude in Sec.~\ref{sec:conclusion}.

ACE is used by different research groups in different contexts and designations, which has led to a family of ACE models and it is sometimes hard to grasp the relation between them. 
For example, the performant implementation of ACE in LAMMPS \cite{Thompson2022LAMMPS} is called PACE\cite{Lysogorskiy2021PACE} and PACEmaker\cite{Bochkarev2022PACEmaker} is software for the parametrization of ACE. PACEmaker enables non-linear and linear models as well as radial basis optimization. Software for obtaining linear ACE models is available as ACEsuit \cite{witt2023ACEjl} and FitSNAP \cite{Rohskopf2023FitSNAP}. 
ACE-related representations for other Lie groups than the rotation group or for global expansions were termed G-equivariant CE \cite{batatia2023general}  or boost-invariant polynomials (BIP) \cite{Munoz2022}, respectively, and it is tempting to call the tensor reduction of ACE coefficients\cite{Darby23} trACE. Message-passing multilayer ACE was abbreviated as multi-ACE \cite{Batatia2022design}, its predecessor with scalar messages was called ml-ACE \cite{Bochkarev2022mlACE} and today's leading implementation of multi-ACE was designated as MACE\cite{batatia2023mace}. The abbreviation of multi-ACE in the MACE implementation should not be confused with magnetic ACE \cite{Rinaldi2023noncollinear}.
Finally, it seems natural to refer to graph ACE as grACE\footnote{The abbreviation GRACE was suggested and used independently also by Cheuk Hin Ho and Christoph Ortner as well as Matous Mrovec.}.

In this paper we show how several of the different ACE variants are comprised in a single, straightforward generalization of ACE. To discriminate between different ACE flavors we specify the model details after the basic name ACE, which we hope contributes to make connections between ACE variants more transparent.

\section{Graph Atomic Cluster Expansion \label{sec:gACE}}

ACE \cite{Drautz19} builds on a decomposition of the energy, or any other scalar, vectorial or tensorial quantity, into atomic contributions $E = \sum_i E_i$, where $i = 1, \dots, N$ indexes the atoms. Such a decomposition is always possible and in itself not an approximation. An approximation is introduced, however, by limiting direct pairwise interactions to a cut-off distance $\rc$, which confines ACE to the local atomic environment of each atom. Here we refer to ACE as local \lACE, in contrast to the global, graph \gACE that we introduce next. For \gACE we do not assume a decomposition into atomic quantities from the very start and  avoiding  this leads us to incorporate semilocal interactions naturally and efficiently. As we will see, local and global ACE are closely related, with \lACE being a subset of \gACE.

We are interested in the energy (or another scalar, vectorial or tensorial property) of a system of $N$ atoms,
\begin{equation}
E = E(\pmb{\sigma}) \,.
\end{equation}
The energy depends on the atomic positions $\br_1, \br_2, \dots, \br_N$, the species of the atoms $\mu_1, \mu_2, \dots, \mu_N$ and potentially other properties such as atomic magnetic moments $\pmb{m}_i$, atomic charges $q_i$, dipole moments, etc. that we do not specify in detail. The ACE framework allows one to incorporate these different dependencies seamlessly in a single coherent model\cite{Drautz2020}. Here we assume that $\mu_i$ collects all atomic variables on atom $i$. The features are combined in the configuration $\pmb{\sigma} = ( \br_1, \mu_1,  \br_2, \mu_2, \dots, \br_N, \mu_N)$ that fully specifies the state of the $N$ atom system.

\subsection{Configurations \label{sec:config}} 
 
In local \lACE, for the evaluation of the energy $E_i$, or other scalar, vectorial or tensorial quantities, the configuration $\pmb{\sigma}$ is centered on atom $i$, $\pmb{\sigma}_i = ( \br_{1i}, \mu_1,  \br_{2i}, \mu_2, \dots, \br_{Ni}, \mu_N)$, with $\br_{ji} = \br_j - \br_i$. The vector $\br_{ii}=0$ of the central atom is ignored, therefore $\pmb{\sigma}_i$ has only $N-1$ entries of position vectors. The configuration is centered on every atom and the energy expressed as $E = \sum_i E_i(\pmb{\sigma}_i)$. Centering the configuration on the atoms brings two advantages, first it facilitates a local expansion of $E_i$ and second, any function of $\sigma_i$ is invariant under translation by construction.

We illustrate atom-centered configurations for four atoms $N=4$. These are given by
\begin{align}
&\pmb{\sigma}_{1} = (\mu_1,  \br_{21}, \mu_2, \br_{31}, \mu_3, \br_{41}, \mu_4) \,, \label{eq:config}\\
&\pmb{\sigma}_{2} = (\br_{12}, \mu_1,   \mu_2, \br_{32}, \mu_3, \br_{42}, \mu_4)\,, \\
&\pmb{\sigma}_{3} = (\br_{13}, \mu_1,  \br_{23}, \mu_2, \mu_3, \br_{43}, \mu_4) \,, \\
&\pmb{\sigma}_{4} = (\br_{14}, \mu_1,  \br_{24}, \mu_2, \br_{43}, \mu_3,  \mu_4) \,.
\end{align}
For $N$ atoms there are $N$ different configurations. These are the configurations that are used in \lACE. 

\begin{figure}
    \includegraphics[width=0.40\textwidth]{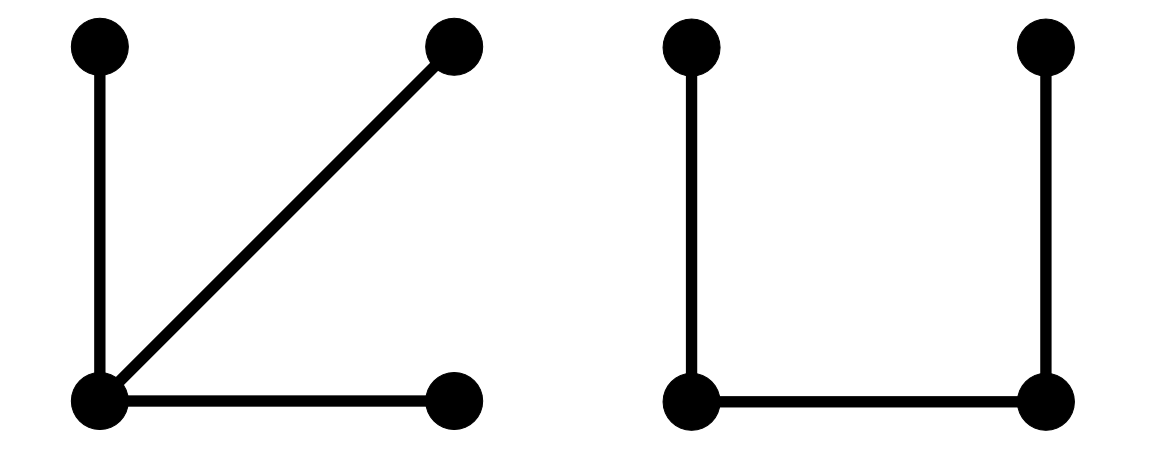}
    \caption{Four atom \lACE configuration (left) and alternative configuration that is added in \gACE (right).}
    \label{fig:configurations}
\end{figure}

There are, however, many additional configurations that provide a full description of atomic positions up to a translation. We give a few examples for four atoms,
\begin{align}
&\pmb{\sigma} = (\mu_1,  \br_{21}, \mu_2, \br_{31}, \mu_3, \br_{42}, \mu_4) \,, \\
&\pmb{\sigma}= (\br_{12},  \mu_1, \mu_2, \br_{31}, \mu_3, \br_{42}, \mu_4) \,, \\
&\pmb{\sigma}= (\mu_1,  \br_{21}, \mu_2, \br_{32}, \mu_3, \br_{42}, \mu_4) \,, \\
&\pmb{\sigma}= (\mu_1,  \br_{21}, \mu_2, \br_{32}, \mu_3, \br_{43}, \mu_4) \,,
\end{align}
etc. or see Fig.~\ref{fig:configurations}. Each of these configurations is complete in a sense that atomic positions can be reconstructed up to a translation. The configurations are not arbitrary, for example, $(\br_{13}, \mu_1, \mu_2, \br_{31}, \mu_3, \br_{42}, \mu_4)$ is not an admissible configuration as it does not allow for a reconstruction of all atomic positions up to a translation.

At first glance the additional configurations appear less attractive than the configurations employed for \lACE, as they are not localized on an atom. As we will see, however, for this reason these configurations are suitable for the description of semilocal interactions. Global, graph \gACE builds on all admissible configurations.

\subsection{Single-particle basis functions \label{sec:singleparticlebasis}}

We attach to each atom $i$ basis functions 
\begin{equation}
\phi_{i u}(\br) = \phi_{i u}(\Delta \br) \,, \label{eq:rbasis}
\end{equation}
with $\Delta \pmb{r} = \br - \br_i$ and where $u$ is a basis function index. Atom-centered basis functions help to ensure translational invariance and introduce a natural, distant dependent interaction hierarchy, but in principle different basis functions are possible, too. We further assume that the basis functions approach zero at some cut-off distance  $\rc$. The cut-off distance can vary from atom to atom, but for ease of notation we just use $\rc$ here. We ask that the basis functions are orthonormal and complete,
\begin{align}
\braket{i u}{j u'} &= \int \phi_{iu}^*(\br) \phi_{ju'}(\br) \, d\Delta \pmb{r} = \delta_{ij}\delta_{uu'} \,, \\
\sum_{iu} \ket{iu}\bra{iu} &= \sum_{iu}  \phi_{iu}^*(\br) \phi_{iu}(\br') = \delta(\br - \br') \,,
\end{align}
Extensions to non-orthogonal basis functions are easily possible as well as to continuous basis functions or different forms of the inner product \cite{Drautz19,Drautz2020}. We further attach to the atoms basis functions $\chi_{i\kappa}(\mu)$ that depend on the species of the atom and/or on other variables. These basis functions can depend on discrete variables, for example, different atomic species or continuous variables such as magnetic moments with magnitudes and directions. We ask these basis functions to be orthonormal and complete on each atom, too,
 \begin{align}
\braket{i \kappa }{i \kappa'} &= \int \chi_{i\kappa}(\mu)^* \chi_{i\kappa'}(\mu) \, d\mu = \delta_{\kappa \kappa'} \,,\label{eq:b21}  \\ 
\sum_{\kappa} \ket{i\kappa}\bra{i\kappa} &= \sum_{\kappa}  \chi_{i\kappa}(\mu)^* \chi_{i\kappa}(\mu') = \delta(\mu - \mu') \,, \label{eq:b22}
\end{align}
 where for discrete variables in Eq.(\ref{eq:b21}) the integration is replaced by a summation and on the right hand side of Eq.(\ref{eq:b22}) the Dirac delta function becomes a Kronecker delta $\delta_{\mu \mu'}$.
  
As we are aiming at representations with well defined transformation under given group operations, we work with basis functions that are also basis functions of the irreducible representations of the group in question, i.e. for the rotation group we use
\begin{equation}
\phi_{i u}(\br) = \phi_{i u}(r,\pmb{e} ) = R_{nl}(r) Y_{lm}\left( \hat{\pmb{r}}\right) \,, \label{eq:sp}
\end{equation}
with the multi-index $u = nlm$, distance dependent radial functions $R_{nl}$ with $r = |\br - \br_i|$ and spherical harmonics $Y_{lm}$ that depend on the direction $\hat{\pmb{r}} = (\br - \br_i)/|\br - \br_i|$. Extension of the formalism to incorporate many other Lie groups with accessible irreducible representations \cite{batatia2023general} or basis functions in Cartesian coordinates \cite{Shapeev2016MTP} $\phi_{i u}(\br) = f(|\br - \br_i|) (\br - \br_i) \otimes (\br - \br_i)  \otimes \dots$ is straightforward \cite{Drautz2020}.

We note that by intuition it should be possible to represent chemistry in terms of distant dependent basis functions of the form Eq.(\ref{eq:rbasis}) only, if these basis functions are made to depend on atomic species. In turn basis functions $\chi_{i\kappa}(\mu)$ that depend on chemistry would not be necessary. This approach is followed, for example, by MTP \cite{Shapeev2016MTP}. We will make the relation to species-dependent basis functions explicit in Sec.~\ref{sec:species}.

\subsection{Cluster basis functions \label{sec:clubasis}}

Next we generate cluster basis functions from products of the single-particle basis functions as,
\begin{equation}
\Phi_{\alpha \vii} = \prod_{k=1}^{N} \phi_{i_k u_k}(\br_k)  \prod_{k'=1}^{N} \chi_{k'} \kappa_{k'}(\mu_{k'}) \,,   \label{eq:clubasis}
\end{equation}
with $1 \leq i_k \leq N$ and $k = 1, \dots, N$. The indices $i_1, i_2 \dots, i_N$ label the configuration $\pmb{\sigma}$ with distance vectors $\br_{k i_k} =\br_{k} - \br_{i_k}$. 
In every configuration exactly one entry fulfills $i_k = k$ and for this $\phi_{k u_k}(\br_k) = 1$, as there are only $N-1$ vectors required to reconstruct atomic positions of $N$ atoms up to a translation.  The products are limited to order $N$ as in a system with $N$ atoms at most $N$ atoms interact.

To aid a local interpretation we ask $\phi_{i 0} = 1$ and $\chi_{i 0}(\mu_i) = 1$. The cluster $\alpha$ collects all atoms $i_k$ with basis function indices $u_k \neq 0$ or $\kappa_{k} \neq 0$. When all basis function indices on all atoms are zero, the cluster $\alpha$ is empty $\alpha = 0$, with $\Phi_{0} =1$. The single-particle basis function indices of cluster $\alpha$ are given by $\vii = (\pmb{u}, \pmb{\kappa})$. The lowest product-order basis functions take the form
\begin{align}
\Phi_{i \vii} &= \chi_{i \kappa}(\mu_i) \,, \\
\Phi_{ij \vii} &= \chi_{i \kappa}(\mu_i)\phi_{i u }(\br_j)   \chi_{j \kappa_1}(\mu_j) \,,\\
\Phi_{ijk \vii} &= \chi_{i \kappa}(\mu_i) \phi_{i u}(\br_j)   \chi_{j \kappa_1}(\mu_j)  \phi_{j u'}(\br_k)   \chi_{k \kappa_2}(\mu_k) \,,  
\end{align}
where these examples show physically and chemically meaningful basis functions as will be discussed in the next Sec.~\ref{sec:graphredtrans}. 

For a given configuration $\pmb{\sigma}_{i_1 i_2 \dots i_N}$, by construction the cluster basis functions are orthonormal and complete,
\begin{align}
\braket{\alpha \vii} {\beta \vii'} &= \SumInt \Phi_{\alpha \vii}^*( \pmb{\sigma}) \Phi_{\beta \vii'}( \pmb{\sigma}) \,d \pmb{\sigma}  = \delta_{\alpha \beta} \delta_{\vii \vii'} \,, \\
\sum_{ \alpha \vii} \ket{ \alpha \vii}\bra{\alpha \vii} &= \sum_{\alpha \vii} \Phi_{\alpha \vii}^*( \pmb{\sigma})  \Phi_{\alpha \vii} ( \pmb{\sigma}') = \delta( \pmb{\sigma} - \pmb{\sigma}' ) \,, \label{eq:complete}
\end{align}
where the sum and integral are carried out over discrete and continuous configuration variables, respectively. The algebra is identical to local \lACE \cite{Drautz19,Drautz2020} and closely related to cluster expansion in alloy theory \cite{Sanchez84}. Different from local \lACE the basis functions are in general not localized on a single atom only. Completeness Eq.(\ref{eq:complete}) holds for every configuration individually, intuitively as it is possible to reconstruct atomic positions up to a translation from each configuration.

As the expansion is complete, for any configuration the energy or other scalar, vectorial or tensorial properties can be represented as a linear combination of the cluster basis functions
\begin{equation}
E = \sum_{ \alpha \vii}  J_{\alpha \vii}  \Phi_{\alpha \vii} ( \pmb{\sigma}) \,. \label{eq:Eall}
\end{equation}
The expansion coefficients are formally obtained as
\begin{equation}
J_{\alpha \vii} = \braket{\alpha \vii} {E} =  \SumInt \Phi_{\alpha \vii}^*( \pmb{\sigma}) E( \pmb{\sigma}) \, d \pmb{\sigma} \,, \label{eq:J}
\end{equation}
where the integral, or sum for discrete variables, is taken over all variables. However, as a configuration in \gACE is in general not localized on an atom, different from \lACE, Eqs.(\ref{eq:Eall}) and (\ref{eq:J}) should be seen as  formal results. Without further analysis they tell us little about the convergence and effectiveness of the expansion.

In the following we work with all possible configurations simultaneously and not only with a single configuration per atom as in \lACE. As every configuration formally provides a complete basis, working with all possible configurations necessarily leads to a globally overcomplete set of basis functions. However, as we will see, this gives us freedom for physical interpretation and will allow us to select sensitive basis functions, which ultimately leads to more accurate models. In order to prepare for this we analyse the cluster basis functions in more detail in the following.

\subsection{Graph structure and reduction \label{sec:graphredtrans}}

The cluster basis functions Eq.(\ref{eq:clubasis}) are evaluated for a given cluster $\alpha$ from an admissible configuration $\pmb{\sigma}$. The clusters are graphs with directed edges that correspond to bonds $i \to j$ decorated with single particle basis functions $\phi_{iu}(\br_j)$ and where atoms correspond to nodes decorated with basis function $\chi_{i \kappa}(\mu)$. Considerable simplifications and reductions of cluster basis functions are achieved by the following observations.

\subsubsection{Topology}  For a given configuration, the expansion coefficient $J_{\alpha \vii}$ associated to cluster basis function $\Phi_{\alpha \vii}$ is independent of atomic positions, from Eq.(\ref{eq:J}). Therefore only graph topology and for a given topology edge orientation and basis function indices need to be considered for the classification of the cluster basis functions and their expansion coefficients $J_{\alpha \vii}$, while the positions of the nodes only affect the numerical values of the cluster basis functions. (For determining graph topology we ignore edge directions. The reason for this will become clear with the introduction of root nodes.)

\subsubsection{Connected graphs} A cluster basis function can only contribute if it is possible to reach any node in the graph from any other node by a walk along graph edges, irrespective of edge orientations. If a cluster basis function consists of two or more graph fragments that are not connected by at least one graph edge, then the graph fragments can be transported rigidly to infinite distance from each other and this will not change the numerical value of the  cluster basis function. Therefore the corresponding cluster basis function is non-local and not suitable for the description of local properties that arise from the interaction of atoms. See illustration in Fig.~\ref{fig:connectedgraphs} b.

\subsubsection{Edges shorter than cut-off distance} 
\begin{figure}
    \includegraphics[width=0.5\textwidth]{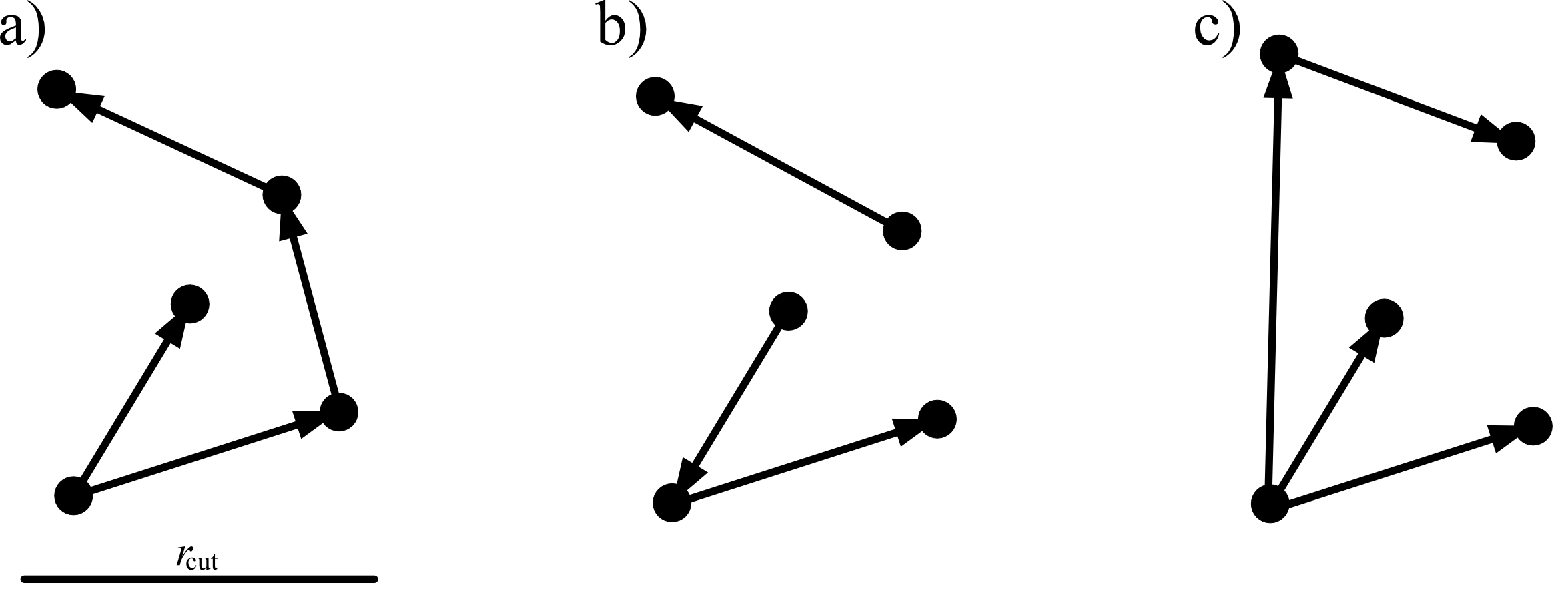}
    \caption{Illustration of locality constraint.  The left hand graph a) is admissible, while the two other graphs are not. The middle graph b) is not a single connected graph, but consists of two disconnected graph fragments. The right hand graph c) includes an edge length larger than cut-off distance.}
    \label{fig:connectedgraphs}
\end{figure}

Cluster basis functions on a graph with an edge $i \to j$  that is longer than $\rc$ vanish identically as $\phi_{iv}(\br_j)=0$. Therefore these cluster basis function cannot contribute, as illustrated in Fig.~\ref{fig:connectedgraphs} c.

\subsubsection{Maximum one incoming edge per node} 
\begin{figure}
    \includegraphics[width=0.45\textwidth]{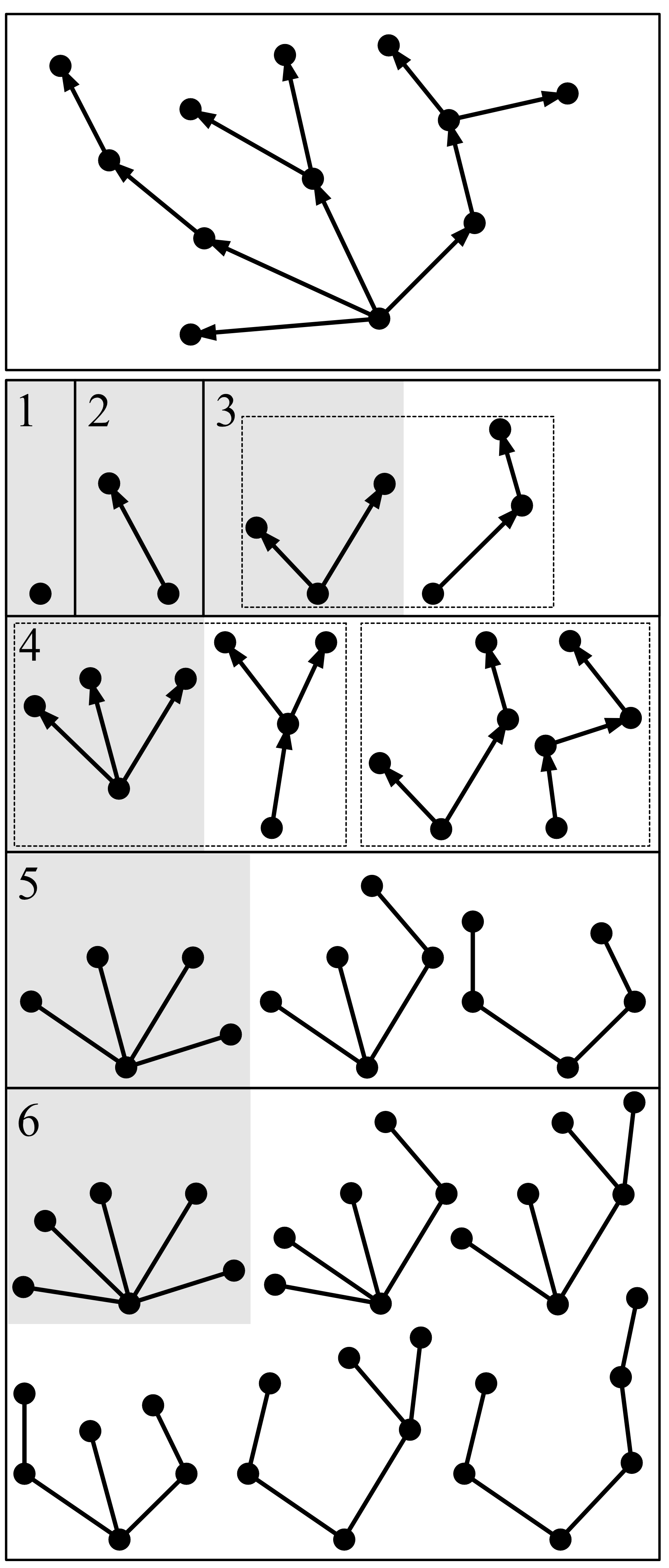}
    \caption{Illustration of an admissible graph with $n-1$ edges and $n$ nodes (top). Possible graphs with up to six nodes (bottom). Star graphs that are part of local \lACE are marked by a grey background. Graphs that have the same topology but different edge orientations are indicated by dashed boxes. For graphs with five and six nodes edge directions are not specified.}
    \label{fig:directedgraphs}
\end{figure}

Up to a maximum of $N-1$ directed edges may come out of a node, while at most a single directed edge may enter any node, as every atomic position is present at most once in any of the cluster basis functions. This limits graphs with $n$ vertices to $n-1$ edges. The graphs are necessarily acylic as a cycle would imply redundant geometrical information that by construction may not be present in admissible configurations. We determine the root node in a graph as the node from which any other node may be reached by directed walks involving edges that point outwards only. In Fig.~\ref{fig:directedgraphs} we illustrate possible graphs and show graphs up to order six. The graphs that are part of local \lACE are stars with all edges attached to a single node.

\subsection{Trees and subtrees \label{sec:trees}}

Admissible graphs are trees. Each of the nodes in a tree can be a root node. As all edges point outwards from the root node, specification of the root node fully determines edge directions, which means that in a tree with $n$ nodes $n$ configurations with different edge orientations exist.
 \begin{figure}[hb!]
    \includegraphics[width=0.4\textwidth]{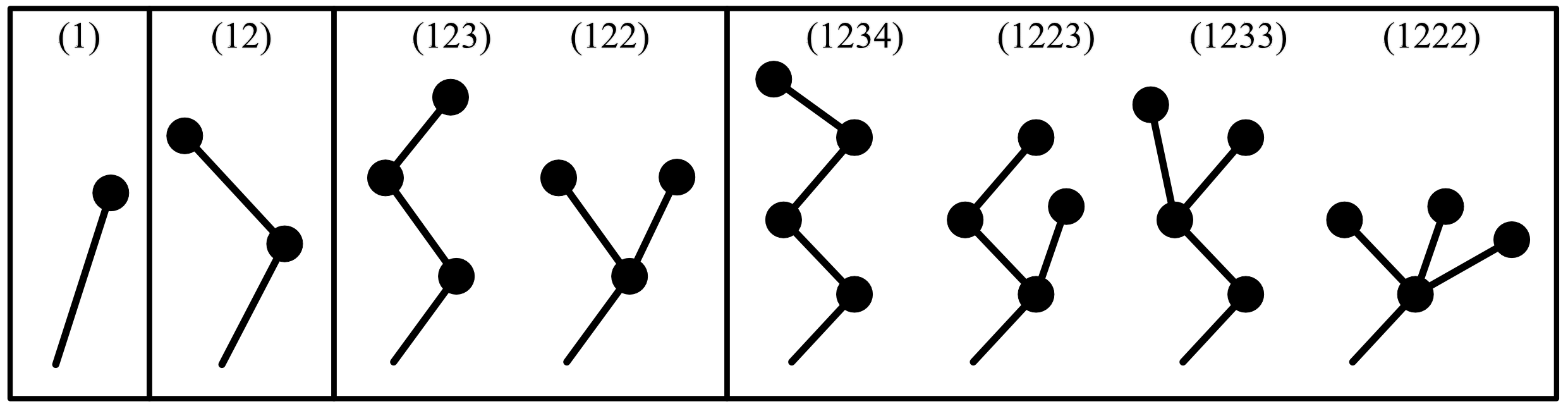}
    \caption{Basic subtrees up to order five including topological classification based on distances from the root node.}
    \label{fig:subtrees}
\end{figure}

Often several subtrees are attached  to a root node. In the following we break trees into subtrees that are attached to the root node. We use a basic topological classification to describe trees and subtrees. For each node the distance from the root node is measured in numbers of edges. The distances are ordered to describe the paths to reach each node from small to large as far as possible, i.e. (123234) is a subtree that branches at node 2 into (23) and (234).  
In Fig.~\ref{fig:subtrees} the topological classification for subtrees with up to five nodes is shown. 

\subsection{Atomic properties \label{sec:root}}

Atoms form the nodes of the graphs of the cluster basis functions. For the computation of properties it seems sensible to associate the contribution of a particular cluster basis function to one or several atoms with the aim of achieving a decomposition into atomic quantities of the form $E = \sum_i E_i$. It is evident that this decomposition is a matter of choice. For example, the contribution  $J_{\alpha \vii}  \Phi_{\alpha \vii}$ of cluster basis function $\Phi_{\alpha \vii}$ could be split equally among the graph nodes. Alternatively, the contribution $J_{\alpha \vii}  \Phi_{\alpha \vii}$ could be added to a particular atom. Formally there are infinitely many ways how to define atomic properties $E_i$ that all lead to the same sum $E = \sum_i E_i$, which also implies that atomic properties $E_i$ are not directly useful for training \cite{Lysogorskiy2023active}. Further constraints are required for a unique definition of $E_i$, such as, for example, symmetrization of many-body interactions and assigning their contributions equally to the nodes \cite{Drautz04}.

Here we choose to evaluate atomic properties by adding the contribution of each graph to its root node. To this end we introduce atomic bases for tree graphs that generalize the atomic base of local \lACE,
\begin{align}
A^{(1)}_{i\kappa v_1} = & \chi_{i \kappa}(\mu_i) \sum_j \phi_{iu_1}(\pmb{r}_{j}) \chi_{j \kappa_1}(\mu_j) =  [ {v_1}  ]_{i\kappa} \,, \label{eq:ab}\\
A^{(12)}_{i \kappa v_1 v_2} = &  \chi_{i \kappa}(\mu_i)  \sum_{j_1 j_2} \phi_{iu_1}(\pmb{r}_{j_1})  \chi_{j_1 \kappa_1}(\mu_{j_1}) \nonumber \\
                      &\times \phi_{j_1 u_2}(\pmb{r}_{j_2})  \chi_{j_2 \kappa_2}(\mu_{j_2}) = [ {v_1}  [ {v_2}  ] ]_{i\kappa} \,, \label{eq:tree12}\\
A^{(11)}_{i \kappa v_1 v_2} = &  \chi_{i \kappa}(\mu_i)  \sum_{j_1 j_2} \phi_{iu_1}(\pmb{r}_{j_1})  \chi_{j_1 \kappa_1}(\mu_{j_1}) \nonumber \\
                      &\times \phi_{i u_2}(\pmb{r}_{j_2})  \chi_{j_2 \kappa_2}(\mu_{j_2}) = [ {v_1}]  [ {v_2}  ]_{i\kappa} \,,  \label{eq:star11}\\  
A^{(123)}_{i \kappa v_1 v_2 v_3} = &  [ {v_1} [ {v_2} [{v_3} ]]]_{i \kappa} \,, \label{eq:tree123}\\
A^{(122)}_{i \kappa v_1 v_2 v_3} = &  [ {v_1} [ {v_2}] [{v_3} ]]_{i \kappa} \,, \\
A^{(112)}_{i \kappa v_1 v_2 v_3} = &  [ {v_1}] [ {v_2} [{v_3} ]]_{i \kappa} \,, \\
A^{(111)}_{i \kappa v_1 v_2 v_3} = &  [ {v_1}] [ {v_2}] [{v_3} ]_{i \kappa} \,, \label{eq:star111}\\
A^{(1234)}_{i \kappa v_1 v_2 v_3 v_4} = &  [ {v_1} [ {v_2} [{v_3} [v_4]]]]_{i \kappa} \,, \\
A^{(1223)}_{i \kappa v_1 v_2 v_3 v_4} = &  [ {v_1} [ {v_2} ] [{v_3} [v_4]]]_{i \kappa} \,, \\
A^{(1222)}_{i \kappa v_1 v_2 v_3 v_4} = &  [ {v_1} [ {v_2} ] [{v_3}] [v_4]]_{i \kappa} \,, \label{eq:ablast}
\end{align}
etc. for higher order, with $v_k= ( u_k, \kappa_k )$ and where the superscript indicates graph topology. The first terms for each order $A^{(1)}$, $A^{(11)}$, $ A^{(111)}$, $ A^{(1111)}$, $\dots$ are the basis functions of local \lACE. The summation over atoms has to be carried out over values $j_1 \neq j_2, j_1 \neq j_3, \dots, j_2 \neq j_3, \dots$ to avoid self-interactions. In Sec.~\ref{sec:self-interaction} we discuss self-interactions that arise from unconstrained summation, i.e. summation that does not respect $j_n \neq j_k$. 

The short-hand notation with square brackets provides a computable form of the graph topology that we use for coding \gACE. Each left square implies advancing across one edge along the tree and summation over neighbors of the following basis function, while closing a square bracket means retreating one edge. The number of left and right square brackets are the same, therefore the root node is always reached at the end. For example, the atomic base of the cluster basis function on tree $(123234)$ with indices $v v_1 v_2 v_3 v_4 v_5$ is computed as $[v[v_1[v_2]][v_3[v_4[v_5]]]]$. 


We can now write the expansion of a property $E$ in terms of atomic contributions $ E = \sum_i E_i$. We order the graphs in the expansion by the number of nodes, corresponding to the body order. Terms including up to five nodes are given as
\begin{widetext}
\begin{align}
&E_i = E_0 + \sum_{\kappa v}  c^{(1)}_{\kappa v}   A^{(1)}_{i\kappa v}  +  \sum_{\kappa v_1 v_2} c^{(11)}_{\kappa v_1 v_2}   A^{(11)}_{i\kappa v_1 v_2} +  \sum_{\kappa v_1 v_2} c^{(12)}_{\kappa v_1 v_2}   A^{(12)}_{i \kappa v_1 v_2}  \nonumber +  \sum_{\kappa v_1 v_2 v_3} c^{(111)}_{\kappa v_1 v_2 v_3} A^{(111)}_{i\kappa v_1 v_2 v_3} + \sum_{\kappa v_1 v_2 v_3} c^{(112)}_{\kappa v_1 v_2 v_3}   A^{(112)}_{i\kappa v_1 v_2 v_3} \nonumber \\
 &+ \sum_{\kappa v_1 v_2 v_3} c^{(122)}_{\kappa v_1 v_2 v_3}   A^{(122)}_{i\kappa v_1 v_2 v_3}  +  \sum_{\kappa v_1 v_2 v_3} c^{(123)}_{\kappa v_1 v_2 v_3}   A^{(123)}_{i\kappa v_1 v_2 v_3}  
 + \sum_{\kappa v_1 v_2 v_3 v_4} c^{(1111)}_{\kappa v_1 v_2 v_3 v_4}   A^{(1111)}_{i\kappa v_1 v_2 v_3 v_4} \nonumber \\ 
& + \sum_{\kappa v_1 v_2 v_3 v_4 } c^{(1112)}_{\kappa v_1 v_2 v_3 v_4}  A^{(1112)}_{i\kappa v_1 v_2 v_3 v_4} + \sum_{\kappa v_1 v_2 v_3 v_4 } c^{(1212)}_{\kappa v_1 v_2 v_3 v_4}   A^{(1212)}_{i\kappa v_1 v_2 v_3 v_4} + \sum_{\kappa v_1 v_2 v_3 v_4 } c^{(1122)}_{\kappa v_1 v_2 v_3 v_4 }  A^{(1122)}_{i\kappa v_1 v_2 v_3 v_4} \nonumber \\
& + \sum_{\kappa v_1 v_2 v_3 v_4 } c^{(1222)}_{\kappa v_1 v_2 v_3 v_4 }  A^{(1222)}_{i\kappa v_1 v_2 v_3 v_4} + \sum_{\kappa v_1 v_2 v_3 v_4 } c^{(1123)}_{\kappa v_1 v_2 v_3 v_4 }  A^{(1123)}_{i\kappa v_1 v_2 v_3 v_4} + \sum_{\kappa v_1 v_2 v_3 v_4 } c^{(1223)}_{\kappa v_1 v_2 v_3 v_4 }  A^{(1223)}_{i\kappa v_1 v_2 v_3 v_4}  \nonumber \\
& + \sum_{\kappa v_1 v_2 v_3 v_4 } c^{(1233)}_{\kappa v_1 v_2 v_3 v_4 }  A^{(1233)}_{i\kappa v_1 v_2 v_3 v_4}  + \sum_{\kappa v_1 v_2 v_3 v_4 } c^{(1234)}_{\kappa v_1 v_2 v_3 v_4 }  A^{(1234)}_{i\kappa v_1 v_2 v_3 v_4} 
+ \dots \label{eq:EgACE}
\end{align}
\end{widetext}
The star graphs $(1)$, $(11)$, $(111)$, $(1111)$ are part of local \lACE, all others are new additions from graph \gACE. 


Clearly global ACE generates a great number of different basis functions and associated coefficients and the main effort in the remainder of this paper is to reduce the complexity of the expansion. Before we get to this, we will discuss transformation under rotation and inversion in the following section.

\section{Equivariance \label{sec:equi}}

For a model of the interatomic interaction we request invariance with respect to translation, rotation, inversion and permutation (TRIP). If the interest is in the expansion of vectorial or tensorial quantities, one requires TRIP equivariance, i.e. rotation of the vectorial output or of the input graphs leads to the same result.
Specifically we require equivariance with respect to the group E(3), which is the semidirect product group of translations T(3) and rotations and inversions O(3), E(3) = T(3) $\rtimes$ O(3).  The orthogonal group O(3) is the group of $3\times 3$ matrices with determinant $\pm1$,  O(3) = C$_2$ $\times$ SO(3), where SO(3) is the group of rotations in three dimensions and C$_2$ the cyclic group of order two for inversion.

Invariance with respect to permutation of identical atoms, i.e. chemical species, is built into \gACE as the cluster basis functions are formed from products of the single particle basis functions, Sec.~\ref{sec:clubasis}. Invariance under translation is ensured as all coordinate inputs are formed as differences that are unchanged by translation. 
To ensure equivariance under rotation and inversion, we rely on the properties of spherical harmonics that are basis functions of the irreducible representations of the rotation group. Evaluation of cluster basis functions on the various graphs leads to products of spherical harmonics. The products of spherical harmonics can be reduced to irreducible representations with generalized Clebsch-Gordan coefficients, as discussed in detail in Refs.~\cite{Drautz2020,Dusson2022,Bochkarev2022PACEmaker}. The mechanism for the reduction to invariant or suitably equivariant basis functions is identical for the star graphs of local \lACE and the tree graphs of global \gACE. Spherical harmonics further possess well defined properties under inversion, which makes it easy to characterize the behavior of the cluster basis functions under inversion.

We therefore follow \lACE to achieve equivariance with respect to rotation and inversion. It suffices to express the expansion coefficients in the form \cite{Drautz2020,Lysogorskiy2021PACE,Dusson2022}
\begin{equation}
{c}_{ \kappa \pmb{\kappa}\pmb{n} \pmb{l} \pmb{m}} = \sum_{\pmb{L} \pmb{M}} \tilde{c}_{ \kappa \pmb{\kappa}\pmb{n} \pmb{l} \pmb{L}} C_{ \pmb{l} \pmb{m}}^{ \pmb{L} \pmb{M}} \,, \label{eq:coeffstruc}
\end{equation}
where $v_i = (\kappa_i n _i l_i m_i)$ are the indices of the cluster basis function, $\pmb{\kappa}\pmb{n} \pmb{l} \pmb{m} = \pmb{v} = (v_1, v_2, \dots)$ and ${c}_{ \kappa \pmb{\kappa}\pmb{n} \pmb{l} \pmb{m}}$ the expansion coefficients in Eq.(\ref{eq:EgACE}). The generalized Clebsch-Gordan coefficients are given by $C_{ \pmb{l} \pmb{m}}^{ \pmb{L} \pmb{M}}$, with angular momenta $\pmb{L}$ for different internal couplings of the coefficients and the expansion coefficients $\tilde{c}_{ \pmb{\kappa}\pmb{n} \pmb{l} \pmb{L}}$ that are independent and trainable parameters. For simplicity we omitted the indices for graph topology, as these are unaffected by the transformation and identical for $c$ and $\tilde{c}$. Note that here we exchanged $c$ and $\tilde{c}$ compared to previous work\cite{Drautz2020,Lysogorskiy2021PACE}, simply in order that we do not have to write expansion coefficients as $\tilde{c}$ everywhere.

The analysis for the group O(3) that trivially extends to E(3) for the simulation of matter in three dimensions is directly transferable to other groups for which irreducible representations and generalized Clebsch-Gordan coefficients are available or can be generated \cite{batatia2023general}, so that \gACE and \lACE are applicable directly in other symmetries and dimensions.

\section{Local and global ACE \label{sec:localglobal}}

The derivation of local \lACE starts by assuming a decomposition of the property of a system into its atomic constituents, $E = \sum_i E_i$. For each of the atomic constituents $E_i$ then an \lACE is carried out. To this end \lACE employs configurations centered on each atom. The derivation of global, graph \gACE is different. \gACE does not decompose a-priori into atomic contributions $E_i$ but provides graph-based cluster basis functions for the complete, global property $E$. By assigning the contribution of the cluster basis functions to their root nodes, the decomposition of $E$ into atomic properties $E_i$ is accomplished only after the formal \gACE construction. Thereby \gACE builds on all possible configurations and not only one, atom-centered configuration per atom as \lACE.

It is evident that local \lACE is a subset of global \gACE. The \lACE is obtained by limiting the \gACE cluster basis functions to stars only. This also implies that for general radial basis functions, \lACE and \gACE are identical up to three-body interactions and differ from four-body interactions onward.

\subsection{Completeness and sensitivity \label{sec:completeness}}

Both local and global ACE expansions are complete by construction. Local \lACE basis functions are complete for each atom, the basis functions suffice to discriminate between all different environments that an atom can have. None of the basis functions are redundant as every basis function contributes information and a description of the local environment of an atom that is orthogonal on all other basis functions.  

In contrast to the local, atomic perspective on completeness of local \lACE, graph \gACE provides an (over)complete set of tree graph-based cluster basis functions for the complete system. \gACE cluster basis functions have $n-1$ edges for $n$ nodes. Edges can connect the same set of nodes in different ways, which results in cluster basis functions with different topology.

\begin{figure}[h!]
    \includegraphics[width=0.37\textwidth]{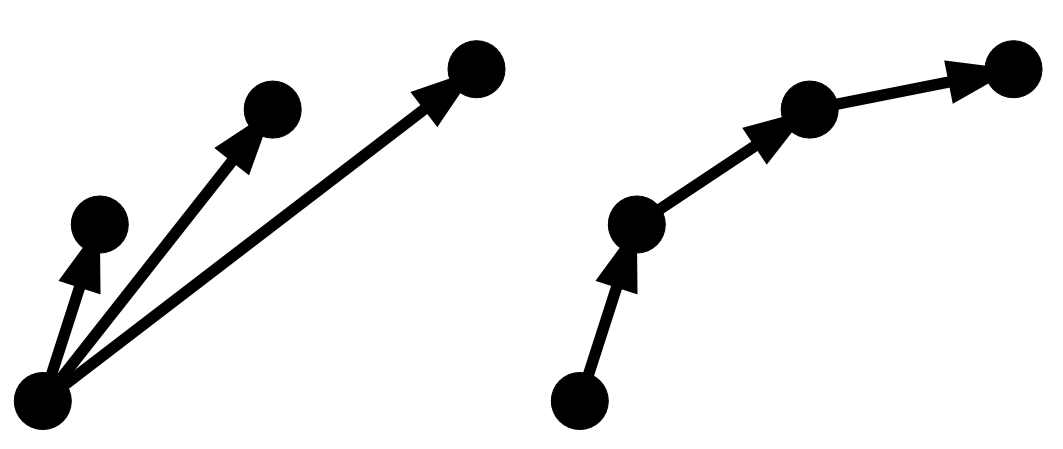}
    \caption{Example of two cluster basis functions on the same nodes but with different topology. The left graph is a star and part of \lACE.}
    \label{fig:examplecomplete}
\end{figure}

Fig.~\ref{fig:examplecomplete} further exemplifies the sensitivity of tree cluster basis functions of \gACE in comparison to stars of \lACE. The narrow opening angles on the root node of the left hand star graph makes it numerically challenging to discriminate the details of atomic positions and one may expect that a relatively large number of spherical harmonics and radial functions are required for an accurate portrayal of the associated four-body interaction. In contrast, the angles between edges in the tree graph on the right-hand side are large and the edge lengths are comparable. One therefore can assume that the tree cluster basis functions are more sensitive to small displacements of the atomic positions than the star cluster basis functions, which in turn should give the tree basis function a greater importance and sensitivity. This argument is further supported by the quantum mechanical analysis in Sec.~\ref{sec:qmfoundation}, where we discuss the role of direct pairwise connectivities for the representation of energies and forces in quantum mechanical systems.

Another obvious distinction between local and global, graph ACE is introduced by the cut-off distance. While \lACE is limited to star graphs with maximum twice the cut-off distance interaction range, the \gACE tree graphs can extend over several cut-off distances. For example, if one assumes that the cut-off distance limits single-particle basis functions to the short pair distances in Fig.~\ref{fig:examplecomplete}, then the contribution of the left-hand star graph must vanish while the right-hand tree graph can contribute. 

\subsection{Self-interactions \label{sec:self-interaction}}

The summation over atoms $j_1, j_2, j_3, j_4,\dots$ in \gACE, Eqs.(\ref{eq:ab})-(\ref{eq:ablast}), is taken for pairwise different values $j_1 \neq j_2, j_1 \neq j_3, \dots, j_2 \neq j_3, \dots$ to avoid self-interactions. For local \lACE the restriction in the summation can be lifted as self-interactions can be accounted for by a renormalization of lower body-order expansion coefficients \cite{Drautz19,Dusson2022}, which in turn enables very efficient implementations~\cite{Lysogorskiy2021PACE}. 

\begin{figure}
    \includegraphics[width=0.50\textwidth]{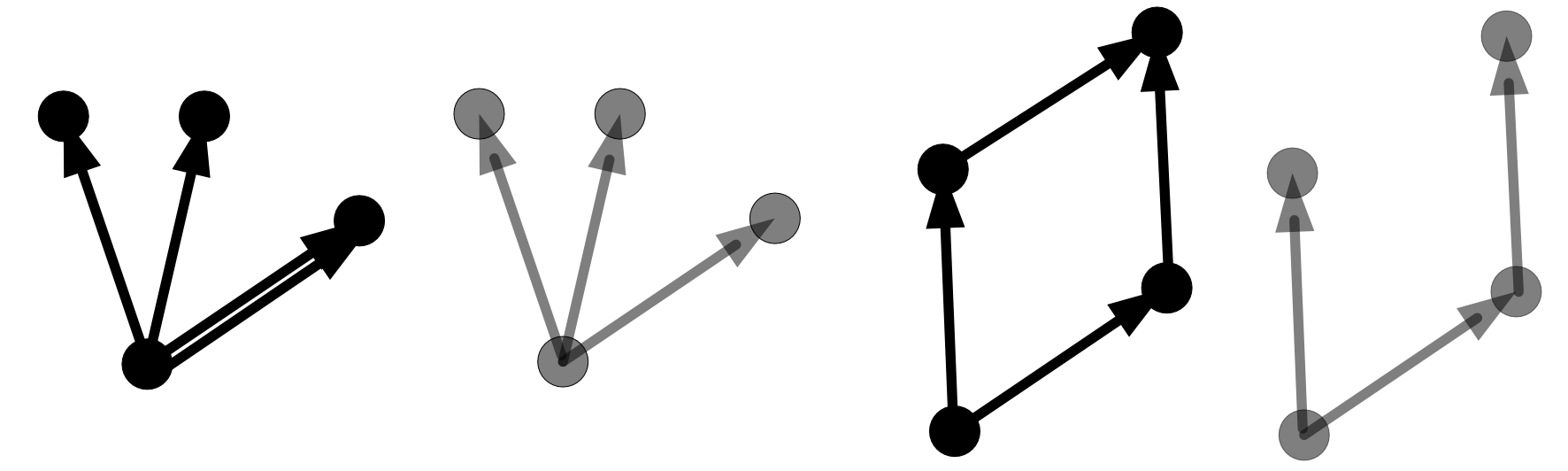}
    \caption{Example of two graphs with self-interacting contributions (black) and two graphs that can contribute to remove these self-interactions (grey).}
    \label{fig:selfinteraction}
\end{figure}

Self-interactions appear to be more difficult to remove in global \gACE as it is not obvious for all graphs that incorporate self-interactions, i.e. when edges connect nodes twice or multiple times, how these can be represented by graphs that are admissible graphs in \gACE as discussed in Sec.~\ref{sec:graphredtrans}. Here we rely on the (over)completeness of the \gACE cluster basis functions, which implies that any graph-based function can indeed be represented by a linear combination of \gACE cluster basis functions. Fig.~\ref{fig:selfinteraction} illustrates self-interacting graphs. On the left hand side a self-interacting star graph is shown together with a star graph with one edge less that enables self-interaction correction. The right-hand side shows a self-interacting graph that is not admissible in \gACE and in grey a graph that can contribute to correcting self-interactions. 

We argue in the next Sec.~\ref{sec:qmfoundation} that some of the self-interactions can be interpreted from quantum-mechanical considerations and can in turn be beneficial for the accuracy of \gACE. We further will see in Sec.~\ref{sec:dandelion} that efficient implementations of \gACE involve self-interactions, just as \lACE, and illustrate this point numerically in Sec.~\ref{sec:4atoms}.

\section{Quantum Mechanical Foundation \label{sec:qmfoundation}}

Graph \gACE closely resembles quantum mechanical electronic structure models, which helps to explain why \gACE is able to model reference data from electronic structure calculations very well. However, as \gACE is not conceived as a quantum mechanical model, not all of its contributions have a direct counterpart in quantum mechanics.  

To make the connection to quantum mechanics explicit, we first simplify to the two-center approximation~\cite{Slater54}, with local orbitals $\ket{\kappa n l m} = R_{nl}^{(\kappa)} Y_{lm}$  on two atoms $i$ and $j$ oriented along the bond axis $\pmb{r}_{ij} =\pmb{r}_i - \pmb{r}_j$. As other atoms in the environment of the bond do not contribute, one assumes invariance under rotation about the bond axis, which implies that the distance-dependent two-center Hamiltonian matrix elements in bond orientation can be written as
\begin{align}
&\beta_{n_1 l_1 m_1 n_2 l_2 m_2}^{(\kappa_1 \kappa_2)}(r_{ij}) = \braHket{\kappa_1 n_1 l_1 m_1}{\hat{H}}{\kappa_2 n_2 l_2 m_2} \nonumber \\
&= \braHket{\kappa_1 n_1 l_1 m_1}{\hat{H}}{\kappa_2 n_2 l_2 m_1} \delta_{m_1 m_2} \,,
\end{align}
with the bond length $r_{ij}$. One rotates the matrix elements to the global coordinate system as
\begin{equation}
H_{n_1 l_1 m_1 n_2 l_2 m_2}^{(\kappa_1 \kappa_2)} = \sum_{m'_1} J( l_1 m_1 l_2 m_2 m'_1) \beta_{n_1 l_1 m'_1 n_2 l_2 m'_1}^{(\kappa_1 \kappa_2)} \,.
\end{equation} 
Spherical harmonics are rotated with Wigner $D$-matrices $D_{m m'}^{(l)}(\alpha \beta \gamma)$ and as rotation about the bond axis is not required, only two angles $\alpha \beta$ are needed. Therefore the Wigner $D$-matrices reduce to spherical harmonics and the transformation matrix $J( l_1 m_1 l_2 m_2 m'_1)$ can be expressed as a linear combination of spherical harmonics
\begin{equation}
J( l_1 m_1 l_2 m_2 m'_1) = \sum_L C(L l_1 l_2 m_1 m_2 m'_1) Y_L^{m_2-m_1} \,,
\end{equation}
and the same holds for the Hamiltonian in the global coordinate system,
\begin{equation}
H_{n_1 l_1 m_1 n_2 l_2 m_2}^{(\kappa_1 \kappa_2)} = \sum_L   \tilde{\phi}^{(\kappa_1 \kappa_2)}_{ L l_1 m_1 l_2 m_2}(r_{ij})  Y_L^{m_2-m_1} \,.
\end{equation} 
See, for example, Eq.(17) in Ref.~\onlinecite{Sharma79} for an explicit expression of $C(L l_1 l_2 m_1 m_2 m'_1)$. The distance dependent function $\tilde{\phi}(r_{ij})$ collects the different pre-factors of the spherical harmonics. 

This means that the two-center matrix elements can be expressed as linear combinations of spherical harmonics with distance dependent pre-factors, which brings a direct link to \gACE basis functions $\phi_{iv}(j)$, Eq.(\ref{eq:sp}), and one can in general represent the two-center Hamiltonian matrix elements from linear combinations of \gACE basis functions $\phi_{iv}(j)$. Conceptually we could therefore replace the \gACE basis functions $\phi_{iv}(j)$ with two-center Hamiltonian matrix elements (obviously at the cost of a more complex disentangling of angular contributions for rotational covariance), or we can just take the view that the \gACE basis functions $\phi_{iv}(j)$ are indeed Hamiltonian matrix elements. This is the view that we take for the further discussion. However, it should also be noted that not all basis functions $\phi_{iv}(j)$ automatically qualify as Hamiltonian matrix elements. Quantum mechanical operators are Hermitian, $\hat{H} = \hat{H}^\dagger$, which imposes symmetries that are not fulfilled by \gACE basis functions in general. We note in passing that overlap matrices are two-center matrices, too, which further reveals a relation between the \gACE basis functions and the Overlap Matrix descriptor of Li et al.~\cite{Goedecker2016fingerprints}

Several linear scaling Density Functional Theory and Tight-Binding methods such as recursion \cite{Haydock80}, Fermi-Operator Expansion \cite{Goedecker94}, Bond-Order Potentials \cite{Horsfield96,Drautz06} and Kernel Polynomial Method \cite{Silver96} rely directly or indirectly \cite{Seiser13,mceniry2017linearscaling} on the evaluation of the moments of the density of states from the atomic coordinates via the moments theorem \cite{Cyrot-Lackmann67}
\begin{align}
\mu_{i\alpha}^{(M)} &= \braHket{i \alpha}{\hat{H}^M}{i \alpha} \nonumber \\
&= \sum_{j_1 \beta_2 j_2 \beta_2 \dots} H_{i \alpha j _1 \beta_1} H_{j _1 \beta_1 j_2 \beta_2} \dots  H_{j _{M-1} \beta_{M-1} i \alpha} \,,
\end{align}
with the matrix elements $H_{i \alpha j \beta} = \braHket{i \alpha}{\hat{H}}{j \beta}$ of the orthonormal basis functions $\braket{i \alpha}{j \beta} = \delta_{i \alpha, j \beta}$ with orbitals $\alpha$ and $\beta$ that are associated to atoms $i$ and $j$, respectively.

\begin{figure}
    \includegraphics[width=0.50\textwidth]{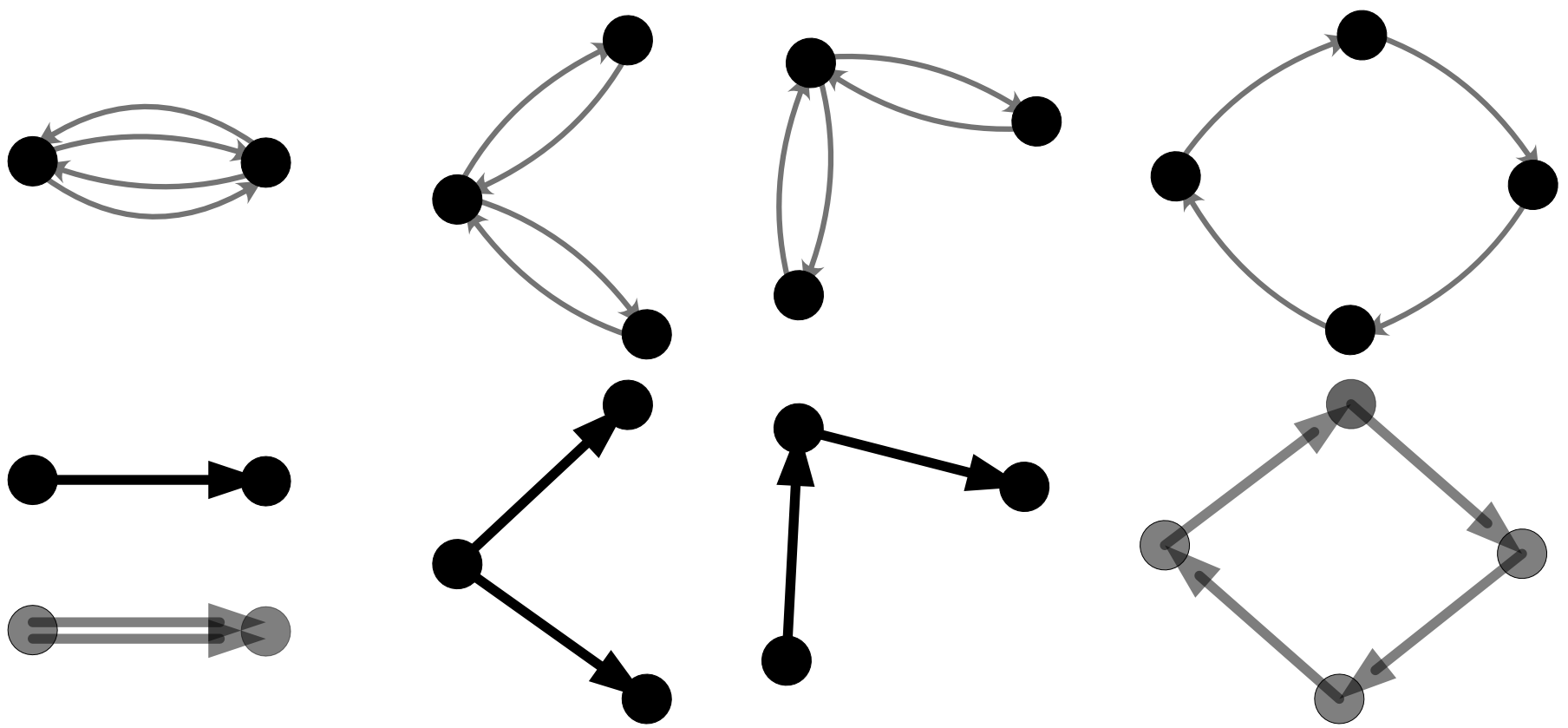}
    \caption{Different contributions to the fourth moment (top row) and their graph representations (bottom row). The rounded arrows represent Hamiltonian matrix elements. On-site matrix elements were ignored. The three-body fourth moment contribution folded onto itself implies self-interactions in \gACE as indicated in grey. The four node cycle graph is self-interacting.}
    \label{fig:4moment}
\end{figure}

Moments of order $M$ of the local density of states of orbital $\alpha$ on atom $i$ are obtained from all self-returning hopping paths, i.e. cycles with $M$ edges, where each edge is given by a Hamiltonian matrix element $H_{i \alpha j \beta}$. Self-interactions in a cycle must be taken into account. Due to self-interactions as well as non-zero onsite matrix elements $H_{i \alpha i \alpha}$ some contributions to the $M$-th moment involve fewer than $M$ atoms. Fig.~\ref{fig:4moment} illustrates contributions to the fourth moment that differ by topology and shows their representation in cluster basis functions. For simplicity on-site matrix elements were ignored, $H_{i \alpha i \alpha} = 0$. Note that the leftmost pair contribution is identical to the self-returning three-body contribution to the right of it, implying a self-interacting pair basis function depicted in grey. 

A detailed analysis of tight-binding models \cite{Pettifor99,Pettifor00,Pettifor02,Drautz04} showed that fourth-moment cycle contributions are small compared to efficient pair and three-body contributions. This is broadly due to different self-cancellation in the graphs in Fig.~\ref{fig:4moment}. The effective pair contribution of the fourth moment has no angular contributions and therefore no cancellation can occur when the sum over neighbors is taken. The effective-three body graph has one opening angle and when the sum is taken over the neighbors of the root atom, angular dependence may lead to some self-cancellation. For the cycle the sum is taken over three neighbors, which due to interfering angular dependencies can lead to effective self-cancellation of the cycle contributions, which render the cycle contribution less important than the pair and three-body contributions to the fourth moment.

\section{Simplifications \label{sec:simple}}

\subsection{Tensor decomposition \label{sec:TD}}

In the following we will discuss possible simplifications of the expression Eq.(\ref{eq:EgACE}). We will assume at various places that tensors can be decomposed as
\begin{equation}
T_{n_1 n_2 n_3 n_4 \dots} = \sum_k \lambda^{(k)} t_{k n_1}^{(1)} t_{k n_2}^{(2)} t_{k n_3}^{(3)} t_{k n_4}^{(4)}  \dots \,, \label{eq:TD1}
\end{equation}
which further implies that decompositions that involve tensors of different sizes are also possible, for example,
\begin{equation}
T_{n_1 n_2 n_3 n_4} = \sum_k \lambda^{(k)} t_{k n_1 n_2 }^{(12)} t_{k n_3}^{(3)} t_{k n_4}^{(4)} \,.\label{eq:TD2}
\end{equation}
Low rank decomposition as provided by a joint summation index $k$ is evidently a key advantage, but not strictly necessary. By trivially expanding $\lambda^{(k)} =  \sum_{k_2 k_3 k_4} \lambda^{(k)} \delta_{ k k_2} \delta_{ k_2 k_3}  \delta_{ k_3 k_4}$ and defining $t^{(l)}_{k_l k_{l+1} n_l} = t^{(l)}_{k_l n_l} \delta_{k_l k_{l+1}}$ from Eq.(\ref{eq:TD1}) one arrives at   
\begin{equation}
T_{n_1 n_2 n_3 n_4 \dots} = \sum_{k_1 k_2 k_3 k_4 \dots} \lambda^{(k_1)} t_{k_1 k_2 n_1}^{(1)} t_{k_2 k_3 n_2}^{(2)} t_{k_3 k_4 n_3}^{(3)} t_{k_4 k_5 n_4}^{(4)}  \dots \,. \label{eq:TD3}
\end{equation}
In the following we decompose cluster basis function expansion coefficients as seems best for an efficient representation of graph \gACE.

Tensor decomposition has been used in the context of ACE recently to eliminate the combinatorial scaling of the number of coefficients with the number of chemical elements by Darby et al.\cite{Darby23}. Here we build on this work, but take a slightly different route. We start by limiting tensor decomposition to chemical indices only. In fact, we deliberately keep the chemical index of the root atom in the associated expansion coefficient for the moment as it appears physically and chemically intuitive to do so. We further discuss in Appendix~\ref{sec:radial} how to completely remove all chemical indices and also employ this reduced representation in our numerical examples.

\subsection{Atomic species \label{sec:species}}

For removing atomic onsite basis functions we limit onsite atomic degrees of freedom to atomic species, i.e., we assume that $\chi_{i\kappa}(\mu_i)$ is a function of the atomic species $\mu_i$ on node $i$ only. The expansion coefficients $c_{\kappa v_1 v_2 v_3 \dots }^{(\dots)}$ of \gACE in Eq.(\ref{eq:EgACE}) with $v = (u, \kappa )$ are rewritten as
\begin{align}
c_{\kappa u_1 \kappa_1 u_2 \kappa_2}^{(t_1 t_2)} &= \sum_k c^{(k)}_{\kappa u_1 u_2} w^{(t_1)}_{k \kappa_1}  w^{(t_2)}_{k \kappa_2} \,, \nonumber \\
c_{\kappa u_1 \kappa_1 u_2 \kappa_2 u_3 \kappa_3}^{(t_1 t_2 t_3)} &= \sum_k c^{(k)}_{\kappa u_1 u_2 u_3} w^{(t_1)}_{k \kappa_1}  w^{(t_2)}_{k \kappa_2}  w^{(t_3)}_{k \kappa_3}  \,, \label{eq:TDchem}
\end{align}
and so on for higher orders. We then modify  basis functions of the form Eq.(\ref{eq:sp}) to have radial functions that depend on chemistry,
\begin{equation}
R_{\mu_j knl}^{(t)}(r_{ji}) = \sum_{\kappa} w^{(t)}_{k \kappa} \chi_{j \kappa}(\mu_j) R_{nl}(r_{ji}) \,. \label{eq:Rde}
\end{equation}
Next, we combine the indices $k$ and $n$ of the radial functions into a joint index $n$ and we do the same for the expansion coefficients $ c^{(k)}_{\kappa v_1 v_2 v_3} \to c_{\kappa v_1 v_2 v_3} $ with $v_i = (n_i l_i m_i)$, which essentially means that we are hiding $k$ in the indices $n_1, n_2, n_3, \dots$. Then, by choosing  $\chi_{i \kappa}(\mu_i) = \delta_{\kappa \mu_i}$ we arrive at simplified expressions for Eq.(\ref{eq:ab}) and following,
\begin{align}
A^{(1)}_{i v_1} =& \sum_j \phi_{iv_1}^{(1)} (\pmb{r}_{j}) =  [ {v}  ]_{i} \,, \label{eq:ab2}\\
A^{(12)}_{i v_1 v_2} =&  \sum_{j_1 j_2} \phi_{iv_1}^{(1)}(\pmb{r}_{j_1})  \phi_{j_1 v_2}^{(2)}(\pmb{r}_{j_2})  = [ {v_1}  [ {v_2}  ] ]_{i} \,,\\
A^{(123)}_{i v_1 v_2 v_3} =&  \sum_{j_1 j_2 j_3} \phi_{i v_1}^{(1)}(\pmb{r}_{j_1}) \phi_{j_1 v_2}^{(2)}(\pmb{r}_{j_2}) \phi_{j_2 v_3}^{(3)}(\pmb{r}_{j_3}) 
\nonumber \\ =&  [ {v_1} [ {v_2} [{v_3} ]]]_{i} \,, 
\end{align}
with the superscript index from the weights $w$ and so on for higher-order and other graphs, and with
\begin{equation}
\phi_{i v}^{(t)}(\br_{j}) = R^{(t)}_{\mu_j nl}(r_{ji}) Y_{lm}\left( \pmb{e}_{ji}\right) \,. \label{eq:sp2}
\end{equation}

We see that the chemistry dependence of the atomic interaction has been fully incorporated into the pairwise, chemistry dependent radial functions and that in this representation the decoration of the nodes with basis functions $\chi_{i\kappa}(\mu_i)$ is not required. In practice we do not need to know the decomposition Eq.(\ref{eq:Rde}) as we optimize the radial functions during training and for this we only need to know the variables $t, \mu_j, n, l$ and $r_{ji}$ that determine the radial functions. Note that the radial functions depend only on the chemical species of the neighboring atom $j$, i.e. $R^{(t)}_{\mu_j nl}(r_{ji})$. From our quantum mechanical considerations in Sec.~\ref{sec:qmfoundation} it may be appropriate to extend this to 
\begin{equation}
\phi_{i v}^{(t)}(\br_{j}) = R_{\mu_j \mu_i nl}^{(t)}(r_{ji}) Y_{lm}\left( \pmb{e}_{ji}\right) \,. \label{eq:sp3}
\end{equation}

We have used radial basis functions of the type $R_{\mu_j \mu_inl}(r_{ji})$ previously \cite{Drautz19,Bochkarev2022PACEmaker}
but ultimately this is a matter of choice and design and not dictated by the formulas that we have derived here.  
Also, as mentioned in Sec.~\ref{sec:singleparticlebasis} the Moment Tensor Potentials~\cite{Shapeev2016MTP} start from chemistry-dependent radial functions and without spanning chemical space with basis functions $\chi_{i\kappa}(\mu_i)$ and we have demonstrated here explicitly that employing chemistry-dependent radial functions or chemical space basis functions can lead to identical representations. 

The expression for atomic properties in \gACE, Eq.(\ref{eq:EgACE}), remains essentially unchanged, but $v$ does no longer contain the chemical index $\kappa$, which means that the number of entries in expansion coefficients $c_{\mu v_1 v_2 v_3}$ in multi-component systems do not suffer from combinatorial explosion. In this way the expansion Eq.(\ref{eq:EgACE}) is written as 
\begin{align}
&E_i = E_0 + \sum_{v}  c^{(1)}_{\mu_i v}    A^{(1)}_{i v} \nonumber \\
 &+  \sum_{v_1 v_2} c^{(11)}_{\mu_i  v_1 v_2}  A^{(11)}_{i v_1 v_2}  +  \sum_{v_1 v_2} c^{(12)}_{\mu_i v_1 v_2}  A^{(12)}_{i v_1 v_2} \nonumber \\
 &+  \sum_{v_1 v_2 v_3} c^{(111)}_{\mu_i v_1 v_2 v_3}   A^{(111)}_{i v_1 v_2 v_3} + \sum_{v_1 v_2 v_3} c^{(112)}_{\mu_i v_1 v_2 v_3}  A^{(112)}_{i v_1 v_2 v_3} \nonumber \\
 &+ \sum_{v_1 v_2 v_3} c^{(122)}_{\mu_i v_1 v_2 v_3}  A^{(122)}_{i v_1 v_2 v_3}+  \sum_{v_1 v_2 v_3} c^{(123)}_{\mu_i v_1 v_2 v_3}  A^{(123)}_{i v_1 v_2 v_3} \nonumber \\
& + \dots \label{eq:EgACE2}
\end{align}
This reduced representation and the full representation  Eq.(\ref{eq:EgACE}) are identical if sufficiently many terms are used in the tensor decomposition Eq.(\ref{eq:TDchem}). However, while the parameterization of multi-component systems is hardly possible with the full representation, the reduced representation enables this.

In the following we further keep Eq.(\ref{eq:EgACE2}) so that it can be read in two ways. If we take the indices as $v_1 = (n_1 l_1 m_1), v_2 = (n_2 l_2 m_2), v_3 = (n_3 l_3 m_3), \dots$, we allow for different radial functions. We can also choose $v_1 = (n l_1 m_1), v_2 = (n l_2 m_2),  v_3 = (n l_3 m_3), \dots$, i.e., a single joint index $n$, as briefly summarized in Appendix~\ref{sec:radial}. It is a matter of choice and numerical considerations, which set of indices is superior and we delay making this choice until the numerical implementation for the examples in Sec.~\ref{sec:application}.

\subsection{Star decomposition and layers \label{sec:star}}

\begin{figure}
    \includegraphics[width=0.5\textwidth]{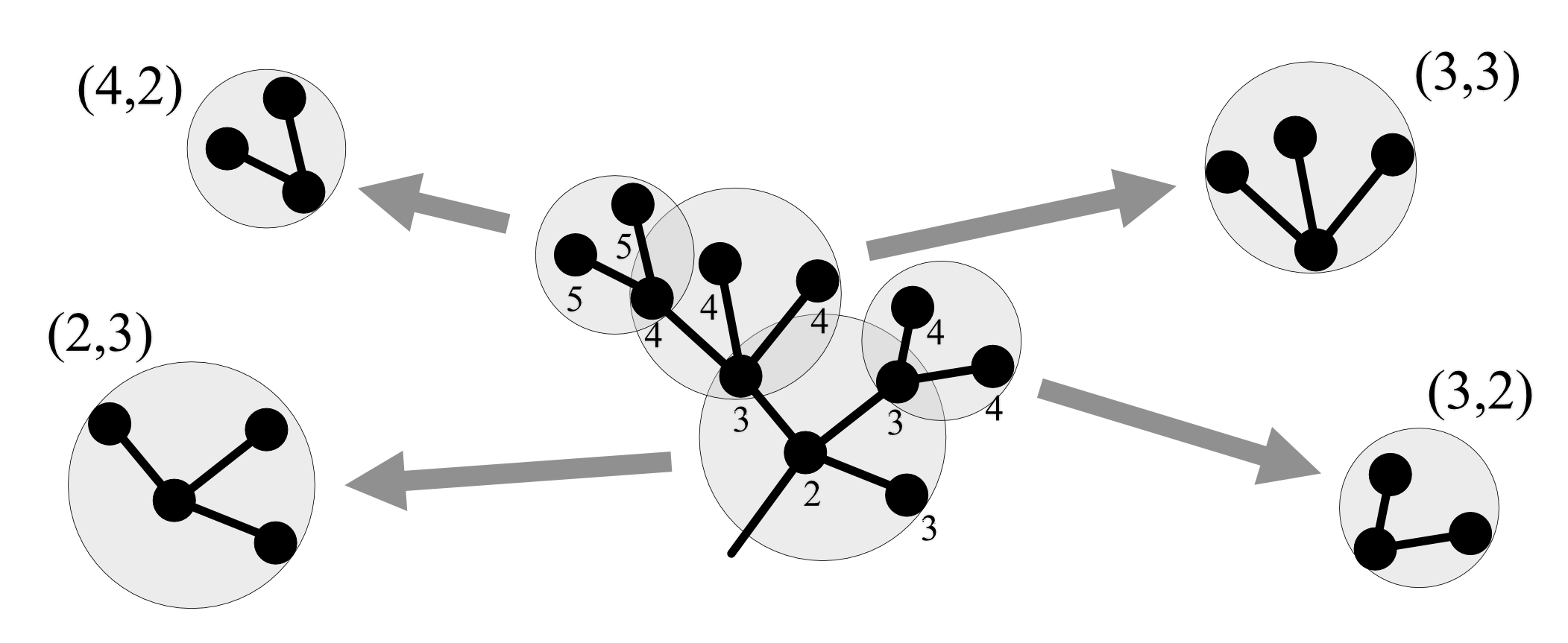}
    \caption{Star decomposition of subtree with topology classification (123344344455). Numbers in the subtree index the distance from the root node. The decomposition is indicated by circles that contain the stars. The stars are classified by their distance from the root node and number of florets.
    }
    \label{fig:startdecomposition}
\end{figure}

We make contact with local ACE$^{(\text{l})}$ by conceptually decomposing trees and subtrees into stars. The stars are categorized by the distance from the root node, i.e. the number of edges required to reach the star along the (sub)tree and the number of outgoing edges of the star. We call the number of edges that are required to reach the star from the root node the \emph{layer} of the star. Star $(t,p)$ is located on the $t$-th layer from the root node and has $p$ outgoing edges. A floret in the following is an outgoing edge together with a node at the outgoing end, where this node has no outgoing edges. See Fig.~\ref{fig:startdecomposition} for an illustration.

We next employ tensor decomposition and write expansion coefficients as products of star expansion coefficients. For example, the expansion coefficient of graph $(12332333)$ is represented as
 \begin{align}
&c_{\mu v_{011} v_{111} v_{112} v_{211} v_{212} v_{221}  v_{222} v_{223} }^{(12332333)} \nonumber \\
&= \sum_k \lambda_k^{(\mu)} c^{(0,1)}_{k v_{011}} c^{(1,2)}_{k v_{111} v_{112}}  c^{(2,2)}_{k v_{211}  v_{212}}   c^{(2,3)}_{k v_{221}  v_{222} v_{223}}\,, \label{eq:TDstar}
\end{align}
where the basis function indices are labeled as $v_{tor}$, with $t$ the layer index, $o$ the index of the star node in layer $t$, $r$ the index of the floret in star $o$. Generalization to arbitrary tree structures is obvious. This representation is general and does not limit interactions associated with trees and subtrees.  The representation further takes into account the symmetry of the graph as illustrated, for example, for the $(1233233)$ subtree, which must be invariant with respect to exchange of the two $(2,2)$ stars, 
 \begin{align}
&c_{\mu v_{011} v_{111} v_{112} v_{211} v_{212} v_{221}  v_{222}}^{(1233233)} \nonumber \\
& =  \sum_k \lambda_k^{(\mu)} c^{(0,1)}_{k v_{011}} c^{(1,2)}_{k v_{111}v_{112}}  c^{(2,2)}_{k v_{211}  v_{212}}   c^{(2,2)}_{k v_{221}  v_{222} }\,. \label{eq:TDstarsym}
\end{align}
As will be discussed next, the decomposition of the expansion coefficients in products of star coefficients Eq.(\ref{eq:TDstar}) has the advantage that the expansion coefficients of all subgraphs of a graph can immediately be written in analogous form.

\subsection{Floret picking and subgraph expansion coefficients}

Starting from a given tree, one can generate all possible trees that are subgraphs of the starting tree by picking one floret after another. For example, picking a floret in the $(2,3)$ star of the $(12332333)$ tree, results in the  $(1233233)$ tree with one node less. Fig.~\ref{fig:dandeloin12bar3} further illustrates floret picking in a $(12333333)$ tree.

In general when a floret on node $o$ in layer $t$ is removed, the number of edges $p$ in the corresponding star is reduced by one. To a removed floret we assign index $v_{top}=0$ and denote the corresponding expansion coefficient as
\begin{equation}
c^{(t,p-1)}_{k v_{to1} \dots v_{top-1}} =  c^{(t,p)}_{k v_{to1} \dots v_{top-1} 0} \,, 
\end{equation}
i.e. the index 0 signals a removed edge. When two florets are picked
\begin{equation}
c^{(t,p-2)}_{k v_{to1} \dots v_{top-2}} =  c^{(t,p)}_{k v_{to1} \dots v_{top-2}0 0} \,, 
\end{equation}
and so on until all florets of the star on the node were removed
\begin{equation}
c^{(t,0)}_{k} =  c^{(t,p)}_{k 0 \dots 0} \,. 
\end{equation}
For the example graph in the previous section, by picking florets on the (2,3) star we generate interaction coefficients of the subgraphs 
\begin{align}
&c_{\mu v_{011} v_{111} v_{112} v_{211} v_{212} v_{221}  v_{222} v_{223} }^{(12332333)} \nonumber \\
&= \sum_k \lambda_k^{(\mu)} c^{(0,1)}_{k v_{011}} c^{(1,2)}_{k v_{111}v_{112}}  c^{(2,2)}_{k v_{211}  v_{212}}   c^{(2,3)}_{k v_{221}  v_{222} v_{223}}\,, \\
&c_{\mu  v_{011} v_{111} v_{112} v_{211} v_{212} v_{221}  v_{222} }^{(1233233)} \nonumber \\
&= \sum_k \lambda_k^{(\mu)} c^{(0,1)}_{k v_{011}} c^{(1,2)}_{k v_{111}v_{112}}  c^{(2,2)}_{k v_{211}  v_{212}}   c^{(2,2)}_{k v_{221}  v_{222} }\,, \\
&c_{\mu  v_{011} v_{111} v_{112} v_{211} v_{212} v_{221}  }^{(123323)} \nonumber \\
&= \sum_k \lambda_k^{(\mu)} c^{(0,1)}_{k v_{011}} c^{(1,2)}_{k v_{111}v_{112}}  c^{(2,2)}_{k v_{211}  v_{212}}   c^{(2,1)}_{k v_{221}  }\,, \\
&c_{\mu  v_{011} v_{111} v_{112} v_{211} v_{212}   }^{(12233)} \nonumber \\
&= \sum_k \lambda_k^{(\mu)} c^{(0,1)}_{k v_{011}} c^{(1,2)}_{k v_{111}v_{112}}  c^{(2,2)}_{k v_{211}  v_{212}}   c^{(2,0)}_{k}   \,,
\end{align}
A star with zero florets is a floret itself that can be picked for further reducing the tree. Thus if we pick the new floret in the above (12233) graph, we arrive at
\begin{align}
c_{\mu  v_{011} v_{111} v_{211} v_{212}   }^{(1233)} &= \sum_k \lambda_k^{(\mu)} c^{(0,1)}_{k v_{011}} c^{(1,1)}_{k v_{111}}  c^{(2,2)}_{k v_{211}  v_{212}}   c^{(2,0)}_{k}   \,.
\end{align}
If we next pick the two florets on the (2,2) star, the resulting expansion coefficient is given by
\begin{align}
c_{\mu v_{011} v_{111}   }^{(12)} &= \sum_k \lambda_k^{(\mu)} c^{(0,1)}_{k v_{011}} c^{(1,1)}_{k v_{111}}    c^{(2,0)}_{k}  c^{(2,0)}_{k}   \,.
\end{align}
The key observation is that by picking florets in all possible ways all subtrees and their corresponding expansion coefficients can immediately be obtained and represented. We will exploit this next for the recursive evaluation of a tree together with all of its subtrees.

\subsection{Recursive evaluation  \label{sec:recursive}}

Because all expansion coefficients of subtrees of a tree can be expanded as part of the representation of the tree expansion coefficient, efficient recursive evaluation becomes possible. We illustrate this here for a small (122) graph, but extension to more complex graphs is obvious and will be exploited in the following section. 

From Eq.(\ref{eq:EgACE2}) we have for the contribution of a (122) tree, including its (12) and (1) subtrees,
\begin{align}
&E_i = E_0 + \sum_{v}  c^{(1)}_{\mu_i v}    A^{(1)}_{i v_1}  +  \sum_{v_1 v_2} c^{(12)}_{\mu_i  v_1 v_2}  A^{(12)}_{i v_1 v_2}  \nonumber \\ &+  \sum_{v_1 v_2 v_3} c^{(122)}_{\mu_i v_1 v_2 v_3}  A^{(122)}_{i v_1 v_2 v_3} \,,
\end{align}
with expansion coefficients
\begin{align}
c_{\mu_i v_1 v_2 v_3}^{(122)} &= \sum_k \lambda_k^{(\mu_i)} c^{(0,1)}_{k v_1}    c^{(1,2)}_{k v_2 v_3}   \,, \\
c_{\mu_i v_1 v_2}^{(12)} &= \sum_k \lambda_k^{(\mu_i)} c^{(0,1)}_{k v_1}    c^{(1,1)}_{k v_2}   \,, \\
c_{\mu_i v }^{(1)} &= \sum_k \lambda_k^{(\mu_i)} c^{(0,1)}_{k v}    c^{(1,0)}_{k }   \,,
\end{align} 
and basis functions
\begin{align}
A^{(1)}_{i v} =& \sum_j \phi_{iv}^{(1)} (\pmb{r}_{j}) \,, \\
A^{(12)}_{i v_1 v_2} =&  \sum_{j j_2} \phi_{iv_1}^{(1)}(\pmb{r}_{j})  \phi_{j v_2}^{(2)}(\pmb{r}_{j_2})  \,,\\
A^{(122)}_{i v_1 v_2 v_3} =&  \sum_{j j_2 j_3} \phi_{i v_1}^{(1)}(\pmb{r}_{j}) \phi_{j v_2}^{(2)}(\pmb{r}_{j_2}) \phi_{j v_3}^{(2)}(\pmb{r}_{j_3})  \,. 
\end{align}
Here the sum of $j, j_2, j_3$ is unrestricted so that we incorporate self-interactions in the expansion, see the discussions in Sec.~\ref{sec:self-interaction}, \ref{sec:qmfoundation} and \ref{sec:4atoms}.

We next define a local \lACE on layer 1 of the expansion
\begin{equation}
\varphi^{(1)}_{jk} =  c^{(1,0)}_{k }  +  \sum_v c^{(1,1)}_{k v}   A^{(2)}_{j v} + \sum_{v_1 v_2}  c^{(1,2)}_{k v_1 v_2}   A^{(2)}_{j v_1}  A^{(2)}_{j v_2} \,, 
\end{equation}
with
\begin{equation}
A^{(2)}_{j v} = \sum_{j_1} \phi_{jv}^{(2)} (\pmb{r}_{j_1}) \,.
\end{equation} 
On the first layer we evaluate
\begin{equation}
A^{(1)}_{i k v} = \sum_{j} \phi_{iv}^{(1)} (\pmb{r}_{j}) \varphi^{(1)}_{jk} \,,
\end{equation} 
and another local \lACE
\begin{equation}
\varphi^{(0)}_{ik} =  c^{(0,0)}_{k }  +  \sum_v c^{(0,1)}_{k v}   A^{(1)}_{i kv}  \,, 
\end{equation}
from which one can identify
\begin{equation}
E_i = \sum_k \lambda_k^{(\mu_i)} \varphi^{(0)}_{ik} \,.
\end{equation}
We see that graph \gACE can be evaluated by layer-wise evaluation of local \lACE and the summation of all channel $k$ contributions at the end. In the following section we will employ this for the evaluation of more complex trees and including all of their subgraph contributions.

\section{Dandelion approximation \label{sec:dandelion}}

\begin{figure}
    \includegraphics[width=0.5\textwidth]{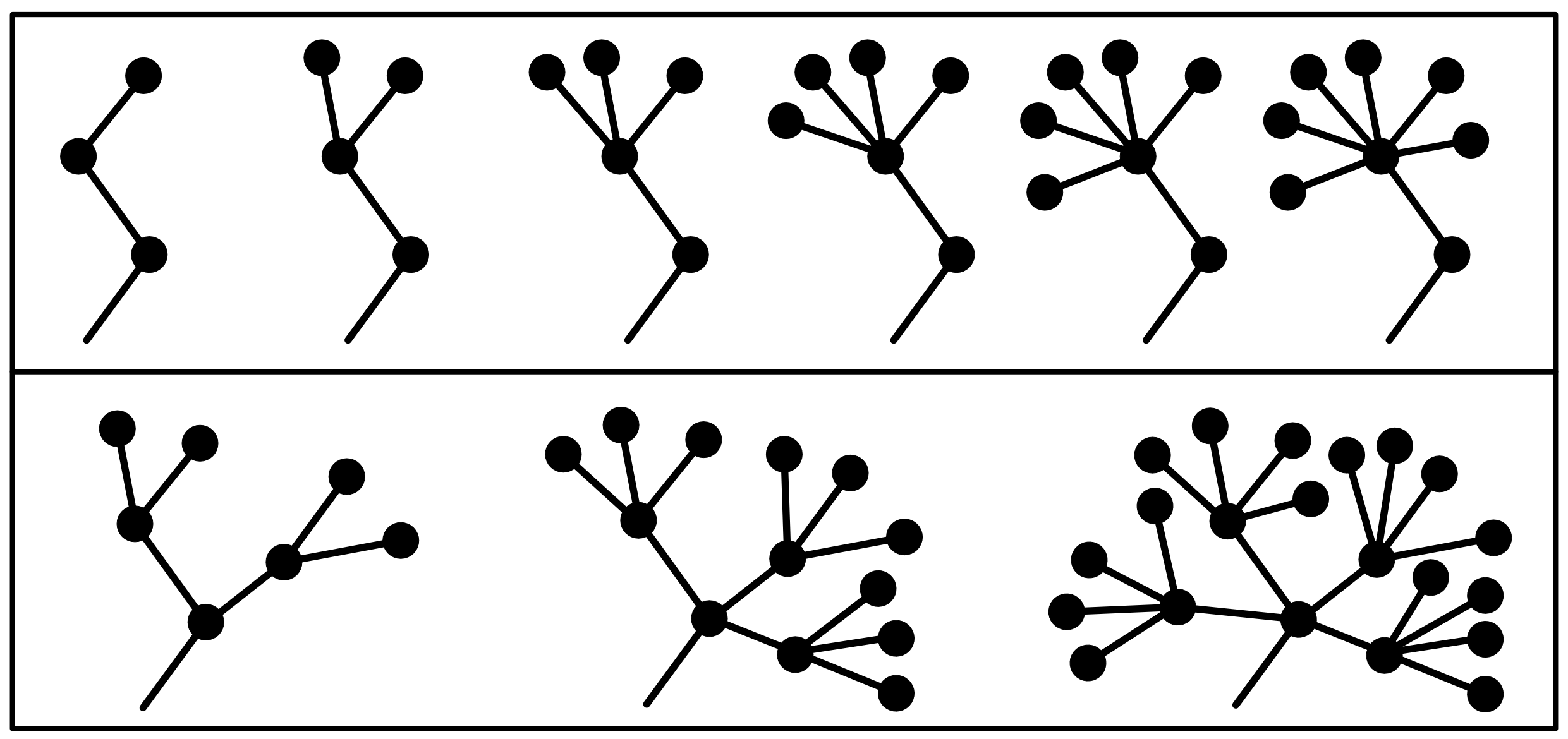}
    \caption{Illustration of $(12\overline{3}_6)$ graph and floret picking (upper panel, from right to left) and  $(1\overline{23}_2)$,  $(1\overline{23}_3)$, $(1\overline{23}_4)$ dandelion graphs (lower panel). The root node is not shown and the subtrees emerge from the root node, which fully characterizes the directions of the edges, therefore edge direction is not shown.}
    \label{fig:dandeloin12bar3}
\end{figure}

In the previous analysis we ordered graphs by their number of nodes. In the following we will drop this hierarchy as this enables us to make use of efficient recursive evaluation of the cluster basis functions. We start by considering trees that are identical up to one node, from which different numbers of edges emerge, i.e. the trees are characterized by topology strings that are identical up to one number that is repeated for a different number of times, for example $(12)$, $(123)$, $(1233)$, $(12333)$, $(123333)$, $(1233333)$ and $(12333333)$. We call the largest of the trees, from which all others can be obtained by floret picking,  a dandelion and abbreviate it as $(12\overline{3}_6)$, see Fig.~\ref{fig:dandeloin12bar3} for an illustration.

We focus on dandelions of the form $(\overline{12 \dots T}_P)$ that have a depth of $T$ layers and every node has $P$ outgoing edges, see lower part of Fig.~\ref{fig:dandeloin12bar3}. The regularity of the dandelion trees makes notation easier and more transparent, while on the other hand the dandelions are sufficiently general to accommodate many other trees as subtrees.

\subsection{Dandelion expansion coefficients and recursion}

We start by decomposing the dandelion expansion coefficient into star contributions following Sec.~\ref{sec:star}. The expansion coefficient of a $(\overline{12 \dots T}_P)$ dandelion can in general be represented as
\begin{align}
c&^{(\overline{12 \dots T}_P)} = \sum_k \lambda_k^{(\mu)} c^{(0,P)}_{k v_{011} \dots v_{01P}} \left( \prod_{p = 1}^P c^{(1,P)}_{k v_{1p1} \dots v_{1pP}} \right) \nonumber \\
& \times \left( \prod_{p = 1}^{P^2} c^{(2,P)}_{k v_{2p1} \dots v_{2pP}} \right) \times \left(\prod_{p = 1}^{P^3} c^{(3,P)}_{k v_{3p1} \dots v_{3pP}} \right) \nonumber \\
& \times \dots \times \left( \prod_{p = 1}^{P^{T-1}} c^{(T-1,P)}_{k v_{T-1,p1} \dots v_{T-1,pP}} \right) \,,
\end{align}
where we did not write out the indices on the left hand side. This formula is best understood for small values and then expanded to general $T$ and $P$. For example, for two layers $T=2$ and and two nodes $P=2$, the expansion coefficient reads
\begin{align}
c^{(\overline{12}_2)} = \sum_k \lambda_k^{(\mu)} c^{(0,2)}_{k v_{011}  v_{012}}  c^{(1,2)}_{k v_{111} v_{112}} c^{(1,2)}_{k v_{121} v_{122}}  \,,
\end{align}
which can be compared directly to the examples given in Sec.~\ref{sec:star}.

Following the discussion in Sec.~\ref{sec:recursive}, the contribution of the $(\overline{12 \dots T}_P)$ dandelion tree and all its subtrees can be evaluated recursively. The recursion is initialized  by a local \lACE in the last layer $T$, with the usual atomic base,
\begin{equation}
A^{(T)}_{i v} = \sum_{j} \phi_{iv}^{(T)} (\pmb{r}_{j}) \,, \label{eq:recini}
\end{equation} 
and \lACE as
\begin{align}
\varphi^{(T-1)}_{jk} &=  c^{(T-1,0)}_{k }  +  \sum_v c^{(T-1,1)}_{k v}   A^{(T)}_{j v} + \sum_{v_1 v_2}  c^{(T-1,2)}_{k v_1 v_2}   A^{(T)}_{j v_1}  A^{(T)}_{j v_2} + \dots \nonumber \\
&+  \sum_{v_1 v_2 \dots v_M}  c^{(T-1,P)}_{k v_1 v_2 \dots v_P}   A^{(T)}_{j v_1}  A^{(T)}_{j v_2} \dots  A^{(T)}_{j L_P}\,. 
\end{align}
The layers $t = T-1, T-2, \dots ,2, 1,0$ are iterated downwards by forming an atomic base that pulls in information from its neighbors in the form of a local \lACE
\begin{equation}
A^{(t)}_{i k v} = \sum_{j} \phi_{iv}^{(t)} (\pmb{r}_{j}) \varphi^{(t)}_{jk} \,.
\end{equation} 
The atomic base is then used to set up an effective \lACE on the next layer,
\begin{align}
\varphi^{(t-1)}_{jk} &=  c^{(t-1,0)}_{k }  +  \sum_v c^{(t-1,1)}_{k v}   A^{(t)}_{j k v} + \sum_{v_1 v_2}  c^{(t-1,2)}_{k v_1 v_2}   A^{(t)}_{j k v_1}  A^{(t)}_{j k v_2} + \dots \nonumber \\
&+  \sum_{v_1 v_2 \dots v_P}  c^{(t-1,P)}_{k v_1 v_2 \dots v_P}   A^{(t)}_{j k v_1}  A^{(t)}_{j k v_2} \dots  A^{(t)}_{j k L_P}\,. \label{eq:layerACEl}
\end{align}
The iteration is terminated with
\begin{equation}
\varphi_i = \sum_k \lambda_k^{(\mu_i)} \varphi^{(0)}_{ik} \,. \label{eq:recfin}
\end{equation}
In practise one may want to stop the \lACE on each layer at a body-order $K$ smaller than $P$, implicitly assuming  $c^{(t,k)} = 0, k = K+1 ,\dots, P$ for the corresponding star expansion coefficients. One can understand the local \lACE on each layer $\varphi^{(t)}_{jk}$ as messages that are attached to the atoms $j$ and transferred to their neighbors $i$ during construction of the effective atomic base $A^{(t)}_{i v}$. 

If desired, the expansion coefficients of the local \lACE on each layer can be tensor-decomposed further, using the approach discussed for \lACE in Sec.~\ref{sec:TD} and Appendix~\ref{sec:radial} or Ref.~\onlinecite{Darby23}.

Further, for a scalar expansion, one can just take \gACE directly as 
\begin{equation}
 E_i = \varphi_i
\end{equation}
or alternatively, in analogy to our work on \lACE \cite{Drautz19,Lysogorskiy2021PACE,Bochkarev2022PACEmaker,Qamar2023}, it can be efficient to compute several expansions for each atom, $\varphi_{i1}, \varphi_{i2}, \varphi_{i3}, \dots$ and compute a scalar property as
\begin{equation}
E_i = \calF(\ace_{i1}, \ace_{i2}, \ace_{i3}, \dots ) \,,\label{eq:EF}
\end{equation}
where $\calF$ is a non-linear function, for example,
\begin{equation}
E_i = \ace_{i1} + \sqrt{\ace_{i2}} \, ,\label{eq:FS}
\end{equation}
for two expansions. 

With each layer in the recursion the effective interaction range is extended by $\rc$. If an atom has $\mathcal{N}$ neighbors within $\rc$, naively one would expect that the effort for evaluating \gACE in dandelion approximation scales as $T^3$ or even $\mathcal{N}^T$. However, as shown in App.~\ref{app:gradients}, where we also provide explicit formulas for gradients, the effort for evaluating \gACE and its gradients scales linearly with the number of layers $T$ and linearly with the number of neighbors $\mathcal{N}$.

\subsection{Explicitly including angular contributions}

We repeat Eqs.(\ref{eq:recini})-(\ref{eq:recfin}) with explicit angular indices. To this end we approximate the reduction of angular products as discussed in Appendix~\ref{sec:pmar}. This approximation allows for different couplings that we illustrate with two examples. 
We write the index $v$ in Eq.(\ref{eq:recini}) and $k$ in Eq.(\ref{eq:layerACEl}) as a combination of angular indices $lm$, parity index $p = \pm 1$ for even and odd parity, respectively, and further radial index $n$ as $nplm$. Eq.(\ref{eq:recini}) then reads
\begin{equation}
A^{(T)}_{i nlm} = \sum_{j} \phi_{inlm}^{(T)} (\pmb{r}_{j})  \,, \label{eq:recinirot}
\end{equation} 
and \lACE, Eq.(\ref{eq:layerACEl}),
\begin{align}
& \varphi^{(T-1)}_{jnplm} =  c^{(T-1,0)}_{nplm }  +  \sum_{n_1l_1m_1} c^{(T-1,1)}_{nplm n_1 l_1 m_1}   A^{(T)}_{j n_1 l_1 m_1} \nonumber \\
&+ \sum_{n_1 l_1 m_1 n_2 l_2 m_2}  c^{(T-1,2)}_{nplm n_1 l_1 m_1 n_2 l_2 m_2}   A^{(T)}_{j n_1 l_1 m_1}  A^{(T)}_{j n_2 l_2 m_2} + \dots \nonumber \\
&+  \sum_{n_1 l_1 m_1 n_2 l_2 m_2 \dots n_P l_P m_P}  c^{(T-1,P)}_{nplm n_1 l_1 m_1 n_2 l_2 m_2 \dots n_P l_P m_P}   \nonumber \\ 
&\times A^{(T)}_{j n_1 l_1 m_1}  A^{(T)}_{j n_2 l_2 m_2} \dots  A^{(T)}_{j n_P l_P m_P}\,. 
\end{align}
The expansion coefficients are represented in the form of Eq.(\ref{eq:coeffstruc}), which corresponds to a transformation into different angular channels $l$. In general the first order term can be simplified as $c^{(T,1)}_{nplm n_1 l_1 m_1} = c^{(T,1)}_{nplm n_1} \delta_{l l_1} \delta_{m m_1}$ as there is no angular momentum coupling possible.

Here we can no longer keep the notation general with respect to radial indices. The representation of the expansion with full tensor decomposition of the radial indices as in Appendix~\ref{sec:radial} is obtained simply by limiting $n = n_1 = n_2 = n_3 = \dots$.

\subsubsection{Independent radial channels}

Next the layers $t = T-1, T-2, \dots, 2, 1, 0$ are iterated downwards. To this end the effective atomic base is computed by pulling in the \lACE from its neighbors. In order to keep tensor dimensions constant during the iteration, we choose to transform the angular character of the basis function $l_1$ and of \lACE, $l_2$, into a joint angular momentum $l$ with Clebsch-Gordan coefficients $C$, 
\begin{align}
&A^{(t)}_{i n n_1 plm} = \sum_{l_1 m_1 l_2 m_2} C_{l_1 m_1 l_2 m_2}^{lm}  \sum_{j} \phi_{i n_1 l_1 m_1 }^{(t)} (\pmb{r}_{j}) \varphi^{(t)}_{jnp_2 l_2 m_2} \,, \label{eq:effArot}
\end{align}
with parity $p = p_2 (-1)^{l_1}$. Here the radial indices $n$ and $n_1$ are not mixed for the construction of the atomic base, but kept independent from the other channels. From the effective atomic base the effective \lACE is computed 
\begin{align}
& \varphi^{(t-1)}_{jnplm} =  c^{(t-1,0)}_{nplm }  +  \sum_{n_1l_1m_1} c^{(t-1,1)}_{nplm p_1 n_1 l_1 m_1}   A^{(t)}_{j n n_1 p_1 l_1 m_1} \nonumber \\
&+ \sum_{n_1 p_1 l_1 m_1 n_2 p_2 l_2 m_2}  c^{(t-1,2)}_{nplm n_1 p_1 l_1 m_1 n_2 p_2 l_2 m_2}   \nonumber \\
&\times A^{(t)}_{j n n_1 p_1 l_1 m_1}  A^{(t)}_{j n n_2 p_2 l_2 m_2} + \dots \nonumber \\
&+  \sum_{n_1 p_1 l_1 m_1 \dots n_P p_P l_P m_P}  c^{(t-1,P)}_{nplm n_1 p_1 l_1 m_1 \dots n_P p_P l_P m_P}   \nonumber \\ 
&\times A^{(t)}_{j n n_1 p_1 l_1 m_1}  A^{(t)}_{j n n_2 p_2 l_2 m_2} \dots  A^{(t)}_{j n n_P p_P l_P m_P}\,. \label{eq:effacerot}
\end{align}
The iteration is terminated with
\begin{equation}
\varphi_{i} = \sum_{nplm} \lambda_{nplm}^{(\mu_i)} \varphi^{(0)}_{inplm} \,. \label{eq:recfinrot}
\end{equation}
For an interatomic potential, in layer 1 only contributions $l = m =0$ and even parity $p=1$ are relevant, but for vectorial or tensorial expansions other values of $l$ and $p$ are of interest~\cite{Drautz2020}.

During the iteration implicitly products of (generalized) Clebsch Gordan coefficients are taken. These products could in principle be reduced to remove some linearly dependent functions~\cite{Dusson2022,Bochkarev2022PACEmaker,goff2023permutationadapted}.

\subsubsection{Coupling radial channels}

While angular coupling seems advisable to maintain clean angular momentum channels, an analogous coupling is also possible for the radial and chemical indices. This can be achieved by introducing additional, layer-dependent weights $W^{(t)}$ in Eq.(\ref{eq:effArot}) to read
\begin{align}
A^{(t)}_{i n plm} = 
\sum_{\substack{n_1 n_2 \\ l_1 m_1 l_2 m_2 \\p_1 p_2}} &W^{(t)}_{\substack{n pl n_1 p_1 l_1 n_2 p_2 l_2}} C_{l_1 m_1 l_2 m_2}^{lm}  \nonumber \\ & \times \sum_{j} \phi_{i n_1 l_1 m_1 }^{(t)} (\pmb{r}_{j}) \varphi^{(t)}_{jn_2 p_2l_2 m_2} \,. \label{eq:effArotalt}
\end{align}

This coupling has the advantage that the effective atomic base has only one radial, respectively, chemical channel as in the original \lACE formalism. The coupling $W^{(t)}$ brings some freedom and its dependency on the indices $n pl n_1 p_1 l_1 n_2 p_2 l_2$ can be explored for numerical efficiency and reduced if necessary. Here full tensor decomposition implies $n = n_1 = n_2$. Other variants, such as requesting $n_1 = n_2$ but keeping $n$ different are also possible and we employ this for the numerical examples in Sec.~\ref{sec:application}.

The effective \lACE is then evaluated as
\begin{align}
& \varphi^{(t-1)}_{jnplm} =  c^{(t-1,0)}_{nplm }  +  \sum_{n_1l_1m_1} c^{(t-1,1)}_{nplm p_1 n_1 l_1 m_1}   A^{(t)}_{j n_1 p_1 l_1 m_1} \nonumber \\
&+ \sum_{p_1 n_1 l_1 m_1 p_2 n_2 l_2 m_2}  c^{(t-1,2)}_{nplm p_1 n_1 l_1 m_1 p_2 n_2 l_2 m_2}   \nonumber \\
&\times A^{(t)}_{j n_1 p_1 l_1 m_1}  A^{(t)}_{j n_2 p_2 l_2 m_2} + \dots \nonumber \\
&+  \sum_{n_1 p_1 l_1 m_1 \dots n_P p_P l_P m_P}  c^{(t-1,P)}_{nplm n_1 p_1 l_1 m_1 \dots n_P p_P l_P m_P}   \nonumber \\ 
&\times A^{(t)}_{j n_1 p_1 l_1 m_1}  A^{(t)}_{j n_2 p_2 l_2 m_2} \dots  A^{(t)}_{j n_P p_P l_P m_P}\,, \label{eq:effacerotalt}
\end{align}
and the recursion terminated with Eq.(\ref{eq:recfinrot}). 

Eqs.(\ref{eq:effArot},\ref{eq:effacerot}) and Eqs.(\ref{eq:effArotalt},\ref{eq:effacerotalt}) represent two alternative recursion variants that build on slightly different flavours of tensor decomposition discussed in Sec.~\ref{sec:TD}. Both variants allow for an accurate recursive evaluation of \gACE in dandelion approximation. 

We also note that while recent graph and message passing interatomic potentials highlighted the importance of non-scalar, equivariant messages, in the recursive evaluation of \gACE equivariant, vectorial and tensorial intermediate \lACE emerge naturally from the angular character of the coupling in the \gACE basis functions.

\section{Comparison to other methods \label{sec:comparison}}

There are two main approaches that claim unification of message passing networks with atom-centered many-atom expansions \cite{Batatia2022design, Nigam2022unified}. In particular the multi-ACE framework is close to results of our analysis as it builds on \lACE messages. We therefore discuss the multi-ACE framework in some detail first.

\subsection{Multi ACE}

The multi-ACE framework \cite{Batatia2022design} has unified \lACE with message-passing graph networks. Multi-ACE shares many features with the recursive evaluation of the dandelion graph of \gACE in Sec.~\ref{sec:dandelion} if the approximate product reduction of angular contributions (Appendix \ref{sec:pmar}) is employed. In fact, the multi-ACE framework is fully contained in the graph \gACE design space that allows for some freedom due to different choices that one can make when tensor-decomposing the \gACE expansion coefficients. A key difference between multi-ACE and \gACE is, however, that we derived the recursive evaluation of the dandelion approximation starting from the global (over)completeness of graph \gACE, while the multi-ACE development was guided by merging \lACE and NequIP \cite{Batzner2022nequip}. For example, message passing in multi-ACE is understood as a chemically inspired sparsification of a local \lACE with a very long cutoff. As discussed in Sec.~\ref{sec:localglobal}, the cluster basis functions of \lACE are a subset of \gACE, and the graph \gACE cluster basis functions enter the recursive evaluation of the dandelion approximation. This also means that multi-ACE cannot be obtained from local \lACE directly, but only from \gACE. Put this way, \gACE provides the foundation and derivation of multi-ACE. 

Multi-ACE makes a distinction between features $h$ and messages $m$. The messages essentially are identical to local \lACE on each layer in the form of Eq.(\ref{eq:effacerot}). The features are obtained from linear transformation of the messages and are multiplied with the single-particle basis functions.  In dandelion approximation of \gACE the distinction between features and messages is not required explicitly, but combined into the computation of the effective atomic base in Eq.(\ref{eq:effArotalt}) and local \lACE on each layer. Furthermore, multi-ACE has a slightly different way of handling angular channels. It keeps two angular indices for the effective atomic base and does not mix one of the two angular channels, in contrast to the mixing of the angular channels in the effective atomic base for the dandelion approximation. It should be emphasized that both approaches are part of the design space of the dandelion approximation and both make use, implicitly or explicitly, of the approximate product reduction of angular contributions discussed in Sec.~\ref{sec:pmar}.

The multi-ACE manuscript \cite{Batatia2022design} discusses design choices explicitly for several message-passing models and how these models can be understood from a multi-ACE perspective. This analysis immediately carries over to graph \gACE and we will therefore limit our discussion to selected representatives of the multi-ACE framework.  

\subsection{Nequip}

NequIP \cite{Batzner2022nequip} pre-dates multi-ACE. It builds on tensor field networks \cite{Thomas2018TensorField} for an accurate representation of the atomic interaction. In retrospect NequIP may be understood as a specific graph \gACE realization, limited to low body-order \lACE messages. Alternatively multi-ACE may be understood as a generalization of NequIP to higher body-order \lACE messages. The limitation to low body-order messages helps to explain why NequIP requires several layers for accurate representations.

\subsection{Multilayer ACE}

The multilayer ACE, abbreviated ml-ACE, is a multi-layer message passing interatomic potential\cite{Bochkarev2022mlACE}. It is limited to scalar, but non-linear update functions and a subset of multi-ACE and of course \gACE. The ml-ACE was inspired by electronic structure relaxation that induces semilocal interactions beyond the immediate neighbor shell of an atom.

\subsection{MACE}

MACE \cite{batatia2023mace,kovacs2023evaluation} is a version and an accurate implementation of multi-ACE. MACE was built on the e3nn software \cite{Geiger2020e3nn} and made a few specific design choices within the multi-ACE framework. For example, MACE used a rank six tensor for representing radial functions despite not indexing chemistry explicitly. This provides some freedom in the multi-ACE design space but cannot be justified directly from our \gACE analysis. MACE used a multilayer perceptron to represent the radial functions, a choice that we followed in our numerical implementation of \gACE. As in the multi-ACE framework, MACE also made a distinction between features and messages and mixes both contributions from one layer into the features of the next layer, while in \gACE this distinction is not necessary. The \lACE on each layer in MACE is constructed recursively, e.g., Sec.~A3.3 in Ref. [\onlinecite{batatia2023mace}]. The recursion is defined to reduce the number of basis function with many edges. Furthermore, MACE did not incorporate all possible parity couplings. As discussed in Sec.~\ref{sec:application}, we observe that MACE requires about one order of magnitude more parameters than \gACE, which may be attributed to some of these design choices.

\subsection{Allegro}

Allegro \cite{Musaelian2022Allegro} is a many-body potential based on iterated tensor products of equivariant representations. The structure of Allegro has been compared in detail to local \lACE  in Ref.~\onlinecite{Musaelian2022Allegro}. For example, the body order of  \lACE has its analogy in the number of layers in Allegro. We speculate that \gACE limited to the same direct neighbor atoms as Allegro has a similar mechanism for maintaining flexible training functions, while in addition benefiting from other graphs than the stars of  \lACE, see the discussion in Sec.~\ref{sec:completeness}. 

\subsection{Unified atom-centred and message-passing schemes}
Nigam et al. \cite{Nigam2022unified} unified atomic density based descriptors and message passing networks by combining the two into multi-centred atomic density correlations. The atomic density correlations have similarities to the graph cluster basis functions that we introduce here, but appear not to be limited to admissible configurations only and in turn also not to acyclic graphs. 

\section{Applications \label{sec:application}}

We consider three application examples. First, a model system of four-atom Si clusters. The four-atom clusters are sufficiently simple to demonstrate and compare various graph architectures, including direct construction of the \gACE basis functions, Eq.(\ref{eq:ab}) and following. At the same time due to the bonding and structural diversity of the reference data it presents a challenge for any machine learning potential. Second, we consider the frequently used revised MD17~\cite{Chmiela2017machine,Christensen2020,revMD17} dataset of 10 small organic molecules as well as a dataset of a flexible 3BPA molecule~\cite{Kovacs2021linACE}. And third, a large carbon dataset that was recently used for a general purpose \lACE ~\cite{Qamar2023}. From these examples, we assess the general applicability of \gACE, its computational performance and its accuracy compared to other models. Details on training and model parameters are given in Appendix~\ref{sec:fitmodel}.

\subsection{Analysis for four-atom Si clusters \label{sec:4atoms}}

\begin{figure}
    \includegraphics[width=0.5\textwidth]{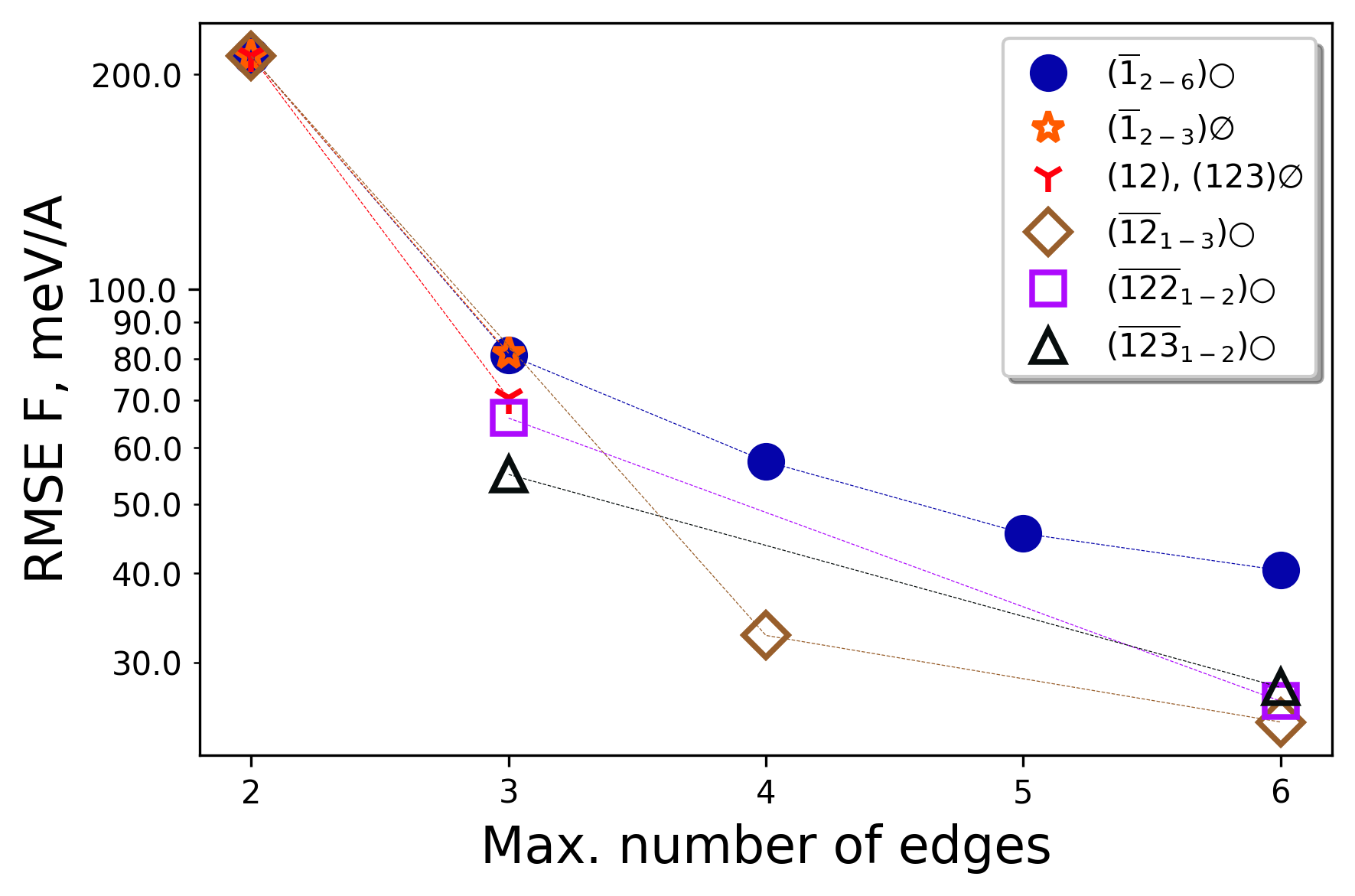}
    \caption{Comparison of different \lACE and \gACE models for four-atom Si clusters with ($\bigcirc$) and without ($\varnothing$) self-interactions. See text for details.
    }
    \label{fig:Siclust3}
\end{figure}

We first consider a dataset of four-atom Si clusters. The dataset consists of a total of 16180 clusters. The clusters were computed with DFT using the PBE functional\cite{Perdew96} with the FHI-aims code \cite{FHIaims,FHIaims2} using tight settings. The calculations were non-magnetic to ensure a single energy hypersurface. The clusters were not relaxed and comprise completely randomized atomic positions. The clusters were constructed such that all distances between atoms are smaller than the cut-off radius of 8 $\mathrm{\AA}$. In this way, we ensure that non of the star graph contributions vanish due to limited interaction range and  instead compare sensitivity of the various graph contributions. From the clusters 90 \% were taken for training and the remaining 10 \% for testing. We compare three different ACE models, local \lACE limited to $(\bar{1})$ stars, an explicit graph basis with star and tree type contributions constructed explicitly and without self-interactions, and global \gACE models in various forms of the dandelion approximation that incorporate self-interactions. Fig.~\ref{fig:Siclust3} shows the test metrics of the considered models as a function of the maximum number of graph edges. For \lACE models the number of edges is the same as the product order. The \lACE results are indicated by filled circles. Star and tree graphs correspond to Eqs.~(\ref{eq:tree12},\ref{eq:star11},\ref{eq:tree123},\ref{eq:star111}). \gACE models were limited to tree graphs and included (12), (1212), (121212) maximum graphs (diamonds), (122) and (122122) maximum graphs (squares), and (123) and (123123) maximum graphs (triangles). As discussed in Sec.~\ref{sec:localglobal}, \lACE and \gACE start to differ only from three edges in the graph. Therefore, different models with two edges, i.e. (11) or (12) graphs, lead to numerically identical results. 

For a maximum of three edges, corresponding to 4-body interactions, as expected the (111) star contribution without self-interactions and the third-order contributions of the atomic base $\left( A^{(1)}\right)^3$, that incorporate self-interactions, are identical. Furthermore, we observe that the (123) tree is more accurate than the (111) star, which highlights the importance of tree-sensitivity as illustrated in Fig.~\ref{fig:examplecomplete}. The dandelion approximation with (122) or (123) largest graphs, which incorporates self-interactions, improves over the explicit (123) graph. The reason for this is that the dandelion recursion brings in additional degrees of freedom, c.f. Eq.(\ref{eq:effArotalt}), due to the approximate product reduction of angular contributions. 
These additional numerical parameters lead to a further lowering of the test error as compared to the (123) tree and, therefore, utilizing these parameters is considered beneficial. 
We note that if Eq.(\ref{eq:effArot}) is used instead, the results for the $A^{(123)}$ dandelion model closely match the (123) tree (not shown).

A maximum of three edges corresponding to four-body interactions should be sufficient for the description of four-atom clusters. In fact, if self-interaction contributions are not permitted, graphs with more than four nodes vanish identically for four-atom clusters. Thus all graphs with four or more edges cannot contribute without self-interactions. However, clearly visible in Fig.~\ref{fig:Siclust3}, increasing the number of edges further lowers the errors for \lACE and \gACE. This is due to the fact that basis functions with up to three edges were not exhaustively increased but kept constant and further self-interacting basis functions with more edges can contribute to lower the error. The rational behind this is our observation that adding a few more basis functions on graphs with additional edges improves the model more efficiently than adding more basis functions to a given graph.

At a maximum of six edges, \gACE improves significantly over \lACE. Thereby the number of layers, i.e. two or three layers, does not appear to be decisive. More accurate models can be achieved by either increasing the number of layers or alternatively the body order of the \lACE in each layer. This explains, for example, why MACE works well with two layers only and many-body \lACE messages while NequIP and Allegro require three layers.


We remark that in Fig.~\ref{fig:Siclust3} we focus on relative errors of different models. Changing hyperparameters, including non-linear embeddings or other forms of the radial functions, would shift the complete graph to lower errors, but without modifying the above analysis and conclusions.

\subsection{Small molecules}

The MD17 dataset~\cite{revMD17} consists of configurations from ab initio molecular dynamics simulations of 10 small molecules. For each molecule, there are 100000 configurations, 1000 of which were randomly selected for training, the rest was used for testing. The graph \gACE mean absolute errors for energies end forces are shown in Table~\ref{table:md17} in comparison to the best available and ACE-related ML models. Results of many other models can be found, for example, in Refs.~\cite{Kovacs2021linACE,batatia2023mace,Batzner2022nequip}. The graph \gACE shows the best performance for all 10 molecules in comparison to all other state-of-the-art models. A further test for which the models were trained only on 50 configurations from the training set is shown on the right side of Table~\ref{table:md17}. Also here \gACE outperforms the other models for every molecule, indicating excellent data efficiency.

The 3BPA dataset was introduced in Ref.~\onlinecite{Kovacs2021linACE} to assess extrapolation capability of machine learned interatomic potentials. The training set consists of 500 configurations of the flexible drug-like molecule 3-(benzyloxy)pyridin-2-amine, obtained from ab initio molecular dynamics at 300~K and the test set contains a series of molecular dynamics calculations at 300~K, 600~K and 1200~K as well as relaxed configurations of so-called dihedral slices that consist of conformer configurations far from the training data. The energy and force RMSE of \gACE for these tests are shown in Table~\ref{table:3bpa} together with results from other state-of-the-art models. \gACE outperforms all other models in all cases. 

\gACE is also computationally more efficient than other methods. The evaluation time of the \gACE models on a NVIDIA A100 GPU is around 2.3 ms/molecule which is about 10 times faster than MACE~\cite{batatia2023mace} and 45 times faster than NequIP~\cite{batatia2023mace} timings reported for the same GPU. Also, models like MACE, Allegro and NequIP utilize several million parameters to achieve the reported performance, while the largest \gACE model uses around 162k parameters. 

\begin{table*}[]
\caption{Energies (E, meV) and forces (F, meV/$\mathrm{\AA}$) MAE of \gACE models for the revised MD17 dataset in comparison to the best published models as well as ACE related models. Values on the left from vertical line show metrics for training on 1000 structures, values on the right are for training on 50 structures. In both cases, metrics are obtained from the 99000 test structures for each molecule. Some of the potentials we reference used a slightly different procedure, for example, for MACE the test set size was 1000 structures only.}
\label{table:md17}
\begin{tabular*}{\linewidth}{@{\extracolsep{\fill}} cccccccc|cccc}
\hline
\hline
 Molecule                       &   & \gACE & ml-ACE\cite{Bochkarev2022mlACE} & MACE\cite{batatia2023mace} & Linear ACE~\cite{Kovacs2021linACE} & NequIP\cite{Batzner2022nequip} & Allegro\cite{Musaelian2022Allegro} & \gACE & Linear ACE & NequIP~\cite{batatia2023mace} & MACE \\ \hline
\multirow{2}{*}{Aspirin}        & E & \textbf{1.7} & 4.7    & 2.2 & 6.1      & 2.3 & 2.3 & \textbf{11.7} & 26.2 & 19.5 & 17.0\\
                                & F & \textbf{6.1} & 14.9   & 6.6 & 17.9     & 8.5 & 7.3 & \textbf{38.2} & 63.8 & 52.0 & 43.9\\ \hline
                                
\multirow{2}{*}{Azobenzene}     & E & \textbf{0.5} & 2.3    & 1.2 & 3.6      & 0.7 & 1.2 & \textbf{4.5} & 9.0 & 6.0 & 5.4 \\
                                & F & \textbf{2.2} & 7.7    & 3.0 & 10.9     & 3.6 & 2.6 & \textbf{16.2} & 28.8 & 20.0 & 17.7 \\ \hline
                                
\multirow{2}{*}{Benzene}        & E & \textbf{0.01} & 0.02   & 0.4 & 0.04     & 0.04& 0.3 & \textbf{0.1} & 0.2 & 0.6 & 0.7 \\
                                & F & \textbf{0.1} & 0.2    & 0.3 & 0.5     & 0.3  & 0.2 & \textbf{1.1} & 2.7 & 2.9 & 2.7 \\ \hline
                                
\multirow{2}{*}{Ethanol}        & E & \textbf{0.2} & 0.7    & 0.4 & 1.2      & 0.4 & 0.4 & \textbf{4.5} & 8.6 & 8.7 & 6.7 \\
                                & F & \textbf{1.4} & 4.4    & 2.1 & 7.3      & 3.4 & 2.1 & \textbf{23.8} & 43.0 & 40.2 & 32.6 \\ \hline
                                
\multirow{2}{*}{Malonaldehyde}  & E & \textbf{0.5} & 1.3    & 0.8 & 1.7      & 0.8 & 0.6 & \textbf{8.4} & 12.8 & 12.7 & 10.0 \\
                                & F & \textbf{3.1} & 8.6    & 4.1 & 11.1     & 5.2 & 3.6 & \textbf{40.9} & 63.5 & 52.5 & 43.3 \\ \hline
                                
\multirow{2}{*}{Naphthalene}    & E & \textbf{0.2} & 0.7    & 0.5 & 0.9      & \textbf{0.2} & 0.5 & \textbf{1.6} & 3.8 & 2.1 & 2.1 \\
                                & F &  \textbf{0.9}  & 3.6   & 1.6 & 5.1      & 1.3 & \textbf{0.9} & \textbf{8.2} & 19.7 & 10.0 & 9.2 \\ \hline
                                
\multirow{2}{*}{Paracetamol}    & E & \textbf{0.9} & 3.2    & 1.3 & 4.0    & 1.4   & 1.5 & \textbf{8.1} & 13.6 & 14.3 & 9.7 \\
                                & F & \textbf{4.0} & 10.7   & 4.8 & 12.7  & 6.9    & 4.9 & \textbf{30.9} & 45.7 & 39.7 & 31.5 \\ \hline
                                
\multirow{2}{*}{Salicylic acid} & E & \textbf{0.5} & 1.5    & 0.9 & 1.8      & 0.7 & 0.9 & \textbf{5.8} & 8.9 & 8.0 & 6.5 \\
                                & F & \textbf{2.7} & 7.7    & 3.1 & 9.3      & 4.0 & 2.9 & \textbf{26.7} & 41.7 & 35.0 & 28.4 \\ \hline
                                
\multirow{2}{*}{Toluene}        & E & \textbf{0.2} & 0.8    & 0.5 & 1.1      & 0.3 & 0.4 & \textbf{2.3}  & 5.3 & 3.3 & 3.1 \\
                                & F & \textbf{1.1} & 4.3    & 1.5 & 6.5      & 1.6 & 1.8 & \textbf{10.9} & 27.1 & 15.1 & 12.1 \\ \hline
                                
\multirow{2}{*}{Uracil}         & E & \textbf{0.2} & 0.6    & 0.5 & 1.1    &  0.4  & 0.6 & \textbf{4.3} & 6.5 & 7.3 & 4.4 \\
                                & F & \textbf{1.7} & 4.0    & 2.1 & 6.6    & 3.2   & 1.8 & \textbf{25.0} & 36.2 & 40.1 & 25.9 \\
\hline
\hline
\end{tabular*}
\end{table*}

\begin{table*}[]
\caption{Energies (E, meV) and forces (F, meV/$\mathrm{\AA}$) RMSE of \gACE for the 3BPA dataset in comparison to other published models. In total five \gACE models were trained, the standard deviation is given in parentheses.}
\label{table:3bpa}
\begin{tabular*}{\linewidth}{@{\extracolsep{\fill}} ccccccc}
\hline
\hline
  Test set                       &   & \gACE           & MACE\cite{batatia2023mace} & NequIP\cite{Musaelian2022Allegro}& BOTNet\cite{Batatia2022design} & Allegro\cite{Musaelian2022Allegro}  \\ \hline
\multirow{2}{*}{300K}            & E & \textbf{1.7}(0.1) & 3.0 (0.5)       & 3.3 (0.1)       & 3.1 (0.1) & 3.8 (0.1) \\
                                 & F & \textbf{7.6}(0.2) & 8.8 (0.3)       & 10.8 (0.2)      & 11.0 (0.1) & 13.0 (0.2) \\ \hline
                                
\multirow{2}{*}{600K}            & E & \textbf{7.3}(0.8) & 9.7 (0.5)       & 11.2 (0.1)      & 11.5 (0.6) & 12.1 (0.5) \\
                                 & F & \textbf{19.2}(0.4)& 21.8 (0.6)      & 26.4 (0.1)      & 26.7 (0.3) & 29.2 (0.2) \\ \hline
                                
\multirow{2}{*}{1200K}           & E & \textbf{25.8}(1.6)& 29.8 (1.0)      & 38.5 (1.6)      & 39.1 (1.1) & 42.6 (1.5) \\
                                 & F & \textbf{61.7}(3.1)& 62.0 (0.7)      & 76.2 (1.1)      & 81.5 (1.5) & 83.0 (1.8) \\ \hline
                                
\multirow{2}{*}{Dihedral Slices} & E & \textbf{7.4}(0.8) & 7.8 (0.6)       & 23.2~\cite{batatia2023mace}& 16.3 (1.5) & - \\
                                 & F & \textbf{15.3}(0.8)& 16.5 (1.7)      & 23.3~\cite{batatia2023mace}& 20.0 (1.2) & - \\ \hline
\hline
\end{tabular*}
\end{table*}

\subsection{Carbon}

\begin{figure}
    \includegraphics[width=0.5\textwidth]{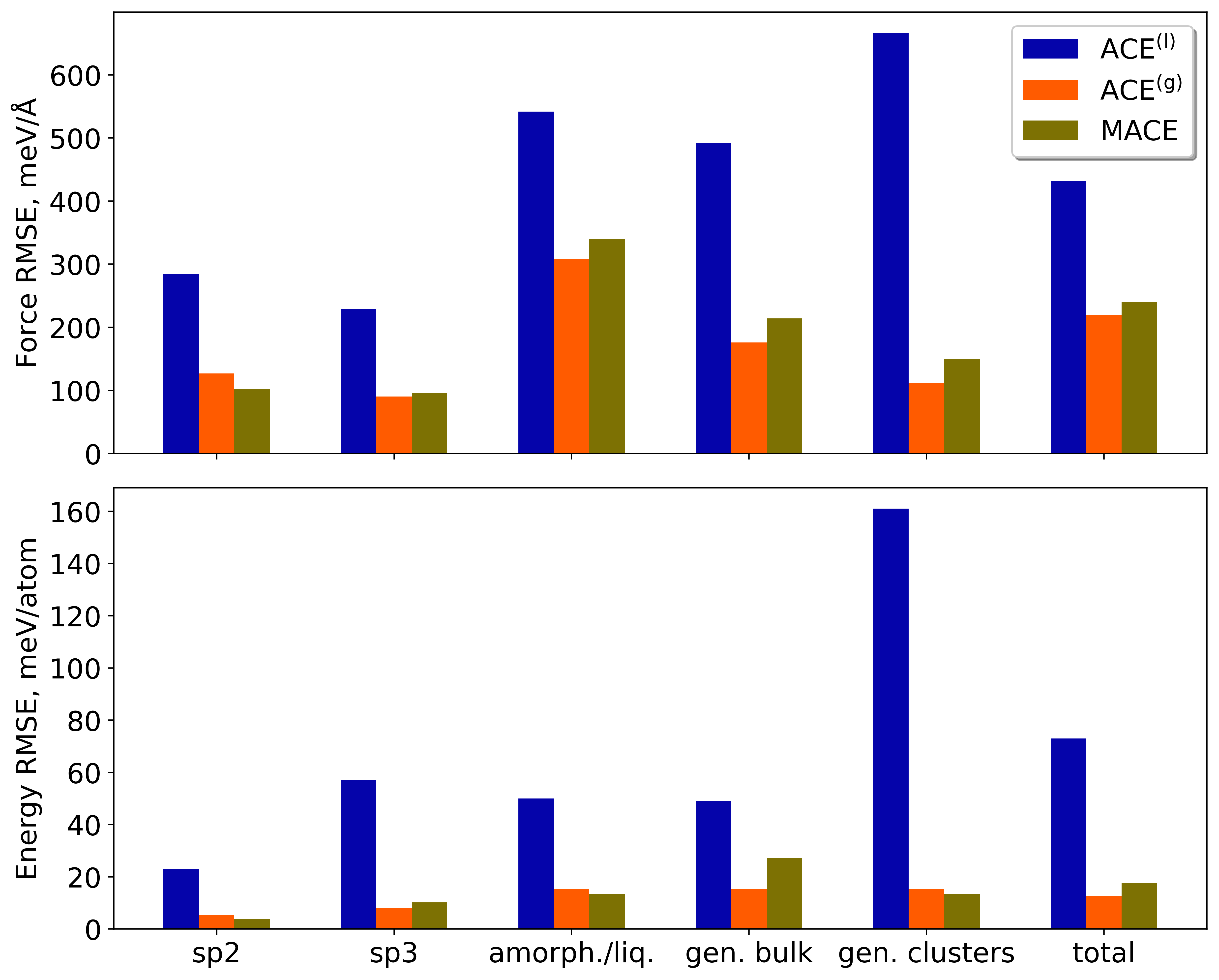}
    \caption{RMSE for energy and forces for different groups of structures in carbon dataset.}
    \label{fig:Cres}
\end{figure}

The carbon dataset~\cite{Qamar2023} is particularly challenging as it was designed for a general purpose potential and therefore consists of structures of various atomic distances and bonding characteristics. The structures can be split into five groups based on bonding character and structure, namely sp2 and sp3 bonded, amorphous and liquid configurations, bulk structures of various crystals and clusters with two to six atoms. The dataset contains in total 19205 structures, and as in the reference, 17293 were taken for fitting and 1912 for testing. We fit \gACE and MACE models and compare to the original \lACE result~\cite{Qamar2023} in Fig.~\ref{fig:Cres}. Both \gACE and MACE provide a considerable improvement over \lACE, especially for clusters. 
While both \gACE and MACE show similar performance with \gACE showing slightly lower errors in total, \gACE is able to achieve this with an order of magnitude fewer parameters, i.e., \gACE has around 105k parameters, while MACE has almost 2.5M. We also assess evaluation time on a NVIDIA A100 GPU for molecular dynamic simulations with LAMMPS. For a supercell with 2016 atoms in equilibrium diamond structure we obtain 200 $\mu$s/atom for \gACE versus 400 $\mu$s/atom for MACE (see~App.\ref{app:compperf} for more details). Moreover, on the GPU with 80 Gb of memory we were able to run simulations only for up to 2016 atoms with the MACE model whereas for \gACE this limit was 12672 atoms.

\section{Conclusion \label{sec:conclusion}}

We introduced graph Atomic Cluster Expansion (grACE or \gACE). Different from \lACE, for the derivation of \gACE we did not assume a decomposition into atomic quantities. Therefore, by construction the \gACE cluster basis functions for each configuration are complete and orthonormal in the space of $N$ interacting atoms. The basis functions from different configurations are categorized by graph topology and their radial, angular, chemical, magnetic, etc. character. A decomposition into atomic contributions is achieved by assigning the contributions of cluster basis functions to the root node of their graphs. 

We show that local \lACE is a subset of global, graph \gACE obtained by limiting \gACE to star graphs only. We further highlight the relation of \gACE to quantum mechanical models. The \gACE basis functions have a close relation to quantum mechanical Hamiltonian product contributions, but do not feature cycle graphs as these are geometrically redundant.

By employing tensor decomposition we achieve an in-princple lossless simplification of \gACE that enables efficient recursive evaluation of tree graphs including the contribution of all subgraphs. In passing this allows us to illustrate that the success of recently developed equivariant message passing models is neither connected directly to message passing nor to equivariance, but should be seen a consequence of including graph basis functions in \gACE that are more sensitive than the stars of \lACE and have a longer reach, which makes them well suited for modelling semilocal interactions.
We show that graph \gACE encompasses multi-ACE and its 
implementation in MACE and 
demonstrate the numerical accuracy and efficiency of graph \gACE in dandelion approximation for molecules, clusters and solids.

In all our tests \gACE is more accurate than the currently most accurate machine learned interatomic potentials, while \gACE is also faster and an order of magnitude more parameter efficient. 

\section*{Acknowledgement}

We acknowledge helpful discussions with Matous Mrovec, Marc Cawkwell, G\'abor Cs\'anyi and Christoph Ortner. AB acknowledges funding by the German Science Foundation (DFG) through CRC 1394.

\appendix

\section{Reducing radial functions \label{sec:radial}}

Radial functions can be further simplified in analogy to reduction of chemical species in Sec.~\ref{sec:species}. To this end we write
\begin{equation}
c_{n_1 l_1 m_1  n_2 l_2 m_2  n_3 l_3 m_3} = \sum_n c^{(n)}_{l_1 m_1 l_2 m_2 l_3 m_3} w^{(1)}_{n n_1}  w^{(2)}_{n n_2}  w^{(3)}_{n n_3}  \,, \label{eq:TDrad}
\end{equation}
which just means that all radial function indices are identical
\begin{align}
A^{(1)}_{i\mu_i n l_1 m_1 } =& \sum_j \phi_{i n l_1 m_1}^{(1)} (\pmb{r}_{j}) =  [ {nl_1m_1}  ]_{i} \,, \label{eq:ab3}\\
A^{(12)}_{i \mu_i n l_1 m_1 l_2 m_2} =&  \sum_{j_1 j_2} \phi_{i n l_1 m_1}^{(1)}(\pmb{r}_{j_1})  \phi_{j_1 n l_2 m_2 }^{(2)}(\pmb{r}_{j_2})  \nonumber \\
=& [ {n l_1 m_1}  [ {n l_2 m_2}  ] ]_{i} \,,\\
A^{(123)}_{i \mu_i n l_1 m_1 l_2 m_2 l_3 m_3} =&  \sum_{j_1 j_2 j_3} \phi_{i n l_1 m_1}^{(1)}(\pmb{r}_{j_1}) \nonumber \\ &\times \phi_{j_1 n l_2 m_2}^{(2)}(\pmb{r}_{j_2}) \phi_{j_2 n l_3 m_3}^{(3)}(\pmb{r}_{j_3}) 
\nonumber \\ =&  [ {n l_1 m_1} [ {n l_2 m_2} [{n l_3 m_3} ]]]_{i} \,, 
\end{align}
with basis functions $\phi_{i n l m}$ given by Eq.(\ref{eq:sp2}). The expression for properties remains analogous to Eq.(\ref{eq:EgACE2}), but with expansion coefficients that are limited to a single radial index $c^{(n)}_{l_1 m_1 l_2 m_2 l_3 m_3}$. See also Ref.~\cite{Darby23}.

\section{Approximate product reduction of angular contributions \label{sec:pmar}}

For invariance and equivariance the expansion coefficients are written in the form of Eq.(\ref{eq:coeffstruc}). This makes tensor decompositions that involve angular contributions more difficult. We decompose an expansion coefficient
\begin{equation}
{c}_{\pmb{n} \pmb{l} \pmb{m}} = \sum_{\pmb{L} \pmb{M}} \tilde{c}_{ \pmb{n} \pmb{l} \pmb{L}} C_{ \pmb{l} \pmb{m}}^{ \pmb{L} \pmb{M}} \,,
\end{equation}
as
\begin{equation}
{c}_{\pmb{n} \pmb{l} \pmb{m}} = \sum_k \lambda_k {c}_{k \pmb{n}_1 \pmb{l}_1 \pmb{m}_1}  {c}_{k \pmb{n}_2 \pmb{l}_2 \pmb{m}_2} \,,
\end{equation}
with $\pmb{n} = (\pmb{n}_1,\pmb{n}_2)$ and analogous for the other indices. We assume that this decomposition is possible in the form of Eq.(\ref{eq:coeffstruc}),
\begin{equation}
{c}_{k \pmb{n}_1 \pmb{l}_1 \pmb{m}_1} = \sum_{\pmb{L}_1 \pmb{M}_1 } \tilde{c}_{k \pmb{n}_1 \pmb{l}_1 \pmb{L}_1 } C_{ \pmb{l}_1 \pmb{m}_1}^{ \pmb{L}_1 \pmb{M}_1} \,,
\end{equation}
and
\begin{equation}
{c}_{k \pmb{n}_2 \pmb{l}_2 \pmb{m}_2} = \sum_{\pmb{L}_2 \pmb{M}_2 } \tilde{c}_{k \pmb{n}_2 \pmb{l}_2 \pmb{L}_2 } C_{ \pmb{l}_2 \pmb{m}_2}^{ \pmb{L}_2 \pmb{M}_2} \,.
\end{equation}
The key is that the generalized Clebsch-Gordan coefficients $C$ do not depend on $k$.

We call this the approximate product angular reduction as a more elaborate analysis must involve the appropriate decomposition of the Clebsch-Gordan coefficients. 

The generalized Clebsch-Gordan coefficients can be generated from products of the Clebsch-Gordan matrices. The products of the Clebsch-Gordan coefficients are overcomplete in a sense that contraction with spherical harmonics in general leads to linearly dependent functions. Therefore not all possible angular momenta couplings in the generalized Clebsch-Gordan coefficients are admissable and selected couplings need to be removed \cite{Bochkarev2022PACEmaker,Dusson2022,goff2023permutationadapted}. Thus if one approximates generalized Clebsch-Gordan coefficients from products of smaller generalized Clebsch-Gordan coefficients, one expects to create too many couplings of which some lead to linearly dependent functions. On the other hand, however, none of the possible angular couplings are missed and in this sense the approximate product angular reduction is overcomplete.

\section{Model and training details \label{sec:fitmodel}}

\subsubsection{Basis functions}

We employ basis functions Eq.(\ref{eq:sp2}) that separate chemistry from radial functions as
\begin{equation}
\phi_{i v}^{(t)}(\br_{j}) = R^{(t)}_{nl}(r_{ji}) W_n(\mu_j) Y_{lm}\left( \pmb{e}_{ji}\right) \,, \label{eq:Rmu}
\end{equation}
with trainable parameters $W_n(\mu_j)$ for differentiating the edge chemistry.

\subsubsection{Radial functions}

Radial functions $R_{nl}(r)$, Eq.(\ref{eq:sp}), in \gACE models are represented as expansions in radial basis functions $g_k (r)$.
We use two types of expansion, namely linear and multi-layer perceptron (MLP). The linear expansion is given by
\begin{equation}
   \label{eq:radialfunc}
	R_{nl}(r) =   \sum_{k} c_{nlk} g_k (r) \,,
\end{equation}
with the radial expansion coefficients $c_{nlk}$. 

For the MLP, a layer transforms inputs $x_k$ to output $h_n$ as
\begin{equation}
   \label{eq:MLPradialfunc}
	h_n =   a\left(\sum_{k} w_{nk} x_k\right) \,,
\end{equation}
where ${a}$ is the activation function and $w_{nk}$ are trainable parameters. 
For the radial function $R_{nl}(r)$, inputs $x_k=g_k (r)$ are transformed via three hidden layers with 64 units each and SiLu activation function. In the last layer no activation is applied. 

Two types of radial basis functions were utilized, the simplified spherical Bessel functions~\cite{Bochkarev2022PACEmaker,Kocer2019} and 
the Bessel function with polynomial envelope function~\cite{Klicpera2020DimeNet,Batzner2022nequip,Musaelian2022Allegro} with $p=5$.

\subsubsection{Non-linear embedding}

The atomic energy is expressed as a non-linear function Eq.(\ref{eq:EF}) with \gACE inputs Eq.(\ref{eq:recfin}). Here we use a single-layer MLP with SiLu activation as a non-linear embedding and vary the number of units in the layer and number of inputs.

\subsubsection{Loss functions}

For optimizing our \gACE models we used the following loss function

\begin{eqnarray}
\label{eq:loss_function}
    \mathcal{L} &=& \kappa_{E}\sum_{n=1}^{N_\mathrm{struct}} w_n^{(E)} \left(E_n^\mathrm{ACE}-E_n^\mathrm{ref}\right)^2  \nonumber \\ 
    &+& \kappa_{F} \sum_{n=1}^{N_\mathrm{struct}} \sum_{i=1}^{n_\mathrm{at,n}} w_{ni}^{(F)} \left(\bm{F}_{ni}^\mathrm{ACE}-\bm{F}_{ni}^\mathrm{ref}\right)^2 \nonumber \\
    &+& \Delta_{L_2}, 
\end{eqnarray}
where $\kappa_E$ and $\kappa_F$ weight the contributions of energies and forces errors, $N_\mathrm{struct}$ is the number of structures employed in the parametrization, and $w_n^{(E)}$ and $ w_n^{(F)}$ are per-structure and per-atom weights for the energy and force contributions for structure $n$, which were set to one for every structure and normalized by the number of structures and force components respectively. For the carbon dataset, the energy residual is in addition normalized to the number of atoms $n_\mathrm{at,n}$ in a structure. $\Delta_{L_2}$ is the regularization term that penalizes the absolute values of the \gACE expansion coefficients $\pmb{c}$ in Eq.(\ref{eq:effacerotalt}) 
\begin{equation}
    \Delta_{L_2} = \kappa_{L_2} \pmb{c}^2\,,
\end{equation}
where $\kappa_{L_2}$ is the regularization weight parameter. 

\subsubsection{Model parameters}

For Si clusters the radial functions are produced from a linear expansion of eight radial basis functions in the from of simplified spherical Bessel functions with cutoff distance of 8$\mathrm{\AA}$ and the radial functions are shared across the layers. We further use $n_{max}=8$ and $l_{max}=4$ for the atomic base on each layer and all intermediate couplings. A linear embedding was used for the energy. For training we used the L-BFGS-B algorithm as implemented in scipy~\cite{2020SciPy-NMeth} with $\kappa_E=1$, $\kappa_F=100$ and $\kappa_{L_2}=5 \cdot 10^{-8}$.

For small molecules, we use models of dandelion type $(\overline{1222}_4)$ with variable radial functions for two layers. A linear expansion of 8 radial basis functions in the from of Bessel functions~\cite{Musaelian2022Allegro} with $n_{max}=8$ is used for the first layer, while a MLP expansion of the same basis is used on the layer zero with $n_{max}=32$. For the MD17 dataset the cutoff for the radial basis is set to 4$\mathrm{\AA}$ and 4.5$\mathrm{\AA}$ for the 3BPA dataset. For all models $l_{max}=3$ for the atomic base and all the basis functions intermediate couplings as well as the second layer \lACE expansion. A single layer MLP with two inputs and 16 hidden units is used for the energy embedding alongside with linear embedding. The L-BFGS-B algorithm was used for training with $\kappa_E=1$, $\kappa_F=500$ and $\kappa_{L_2}=5 \cdot 10^{-5}$ for all the models except benzene for which was $\kappa_F=1000$.

For carbon a $(\overline{1222}_3)$ model was used, with radial functions in the form of a MLP expansion with Bessel functions~\cite{Klicpera2020DimeNet} input, $n_{max}=16$ and cut-off 5$\mathrm{\AA}$, which is shared across layers. We further set $l_{max}=4$ for atomic base and intermediate couplings on the layer zero. The first layer \lACE expansion and basis functions intermediate couplings were set to $l_{max}=3$. We utilize a single layer MLP with 16 inputs and 32 hidden units for the energy embedding together with linear embedding. For training we used the AMSGrad version of the Adam~\cite{kingma2017adam} optimizer with batch size 10, learning rate $5 \cdot 10^{-3}$ and $\kappa_E=1$, $\kappa_F=500$. During optimization, the learning rate was reduced in steps with a factor 0.6 down to $6 \cdot 10^{-4}$. With each reduction the fit was restarted and energy and force weights in the loss functions were gradually adjusted to $\kappa_E=10$, $\kappa_F=1$ as well. Finally, the fit was further refined with the L-BFGS-B algorithm. 

For training the carbon MACE model we use the same train and test split as above, 256 feature channels, $l_{max}=3$, messages with $L_{max}=2$ and radial cutoff of 5$\mathrm{\AA}$. Optimization is performed with the AMSGrad version of the Adam optimizer with learning rate 0.01 and batch size 5. Energy and force weights in the loss function were set to 80 and 1000 respectively followed by switching energy weight to 1000 and force weight to 10.

The \gACE models and fitting algorithm are implemented within tensorpotential package~\cite{Bochkarev2022PACEmaker} and GPU acceleration is enabled via TensorFlow~\cite{tensorflow2015-whitepaper}. All \gACE fits and evaluations were performed with double precision.

\section{Comparison of computational performance}
\label{app:compperf}
\begin{figure}
    \includegraphics[width=0.5\textwidth]{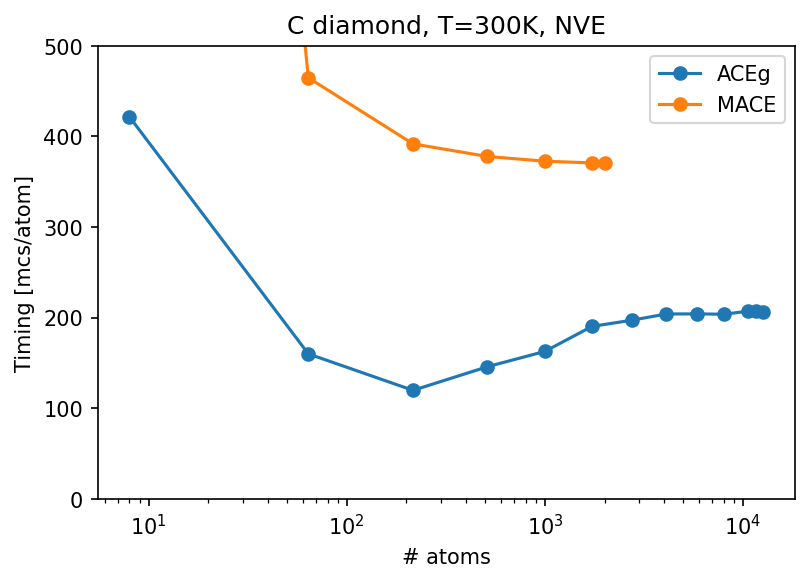}
    \caption{Computational performance for \gACE and MACE models for carbon. See text for details.}
    \label{fig:acegperf}
\end{figure}

A comparison of computational performance for \gACE and MACE models is presented in Fig.~\ref{fig:acegperf}. All simulations were performed using LAMMPS~\cite{Thompson2022LAMMPS} with custom pair styles for \gACE and MACE, without domain decomposition. Evaluation of MACE was carried out exclusively on the GPU, with the KOKKOS package for neighbor list construction and MD trajectory integration. 
In contrast, the current \gACE is implemented in LAMMPS calling the TensorFlow graph on the GPU to compute energies and forces, while all other calculations run on a single CPU core, which requires CPU-GPU data transfer in both directions. Therefore, the current \gACE implementation can be substantially improved in the future.

\section{Efficient gradients for graph ACE \label{app:gradients}}

\gACE graph basis functions extend over $T$ layers, where each layer covers a cut off distance $\rc$. Therefore, in a homogeneous system with $\mathcal{N}$ neighbors around each atom, one can assume that the root atom interacts with $T^3 \mathcal{N}$ atoms and one may expect the effort for evaluating \gACE to scale as $T^3$ with the number of layers. Alternatively, one could argue that in each layer connections are made to each of the $\mathcal{N}$ neighbors and from this expect an even steeper scaling as  $\mathcal{N}^T$. 

Here we show that the evaluation of \gACE, including gradients, can be achieved with an effort that scales linearly with neighbors $\mathcal{N}$ and linearly with layers $T$ only. This is achieved by a straightforward change of summation order, in close analogy to the linear scaling of \lACE with $\mathcal{N}$.

We note that the implementation that we used to carry out the numerical examples and to establish timings presented here exploits the automatic differentiation available from TensorFlow and does not employ the following formulas for gradients directly. The reasons for this is that the optimization of the loss functions requires gradients with respect to positions for the forces and subsequent derivatives of the forces with respect to expansion coefficients. It will be interesting to compare the efficiency of TensorFlow automatic gradients with the explicit implementation of the following gradient for force evaluation.

We give explicit formulas for the case of coupled radial channels, but for other variants of \gACE the derivation is essentially identical. We introduce
\begin{align}
\tilde{\phi}^{(t)}_{ji n plm} &= \sum_{\substack{n_1 n_2 \\ l_1 m_2 l_2 m_2}} W^{(t)} _{n l  n_1 p_1 l_1  n_2 p_2 l_2} \nonumber \\
&\times C_{l_1 m_1 l_2 m_2}^{lm} \phi_{i n_1 p_1 l_1 m_1 }^{(t)} (\pmb{r}_{j}) \varphi^{(t)}_{jn_2 p_2l_2 m_2} \,, \label{eq:tildephi}
\end{align}
with $p_1 = (-1)^{l_1}$ and $p = p_1 p_2$. Leaving away all but the atomic and layer indices and writing $ \phi_{ji}^{(t)} =\phi_{i}^{(t)} (\pmb{r}_{j})$, this reads
\begin{equation}
\tilde{\phi}_{ji}^{(t)} = W C \phi_{ji}^{(t)} \varphi^{(t)}_{j} \,.
\end{equation}
The atomic base is computed as
\begin{equation}
A_{i}^{(t)} = \sum_j^{\mathcal{N}_i} \tilde{\phi}_{ji}^{(t)} \,,
\end{equation}
where $\mathcal{N}_i$ contains all atoms within the cut-off distance $\rc$ around atom $i$ but not atom $i$ itself. Local \lACE on the next layer is a polynomial of the atomic base
\begin{equation}
\varphi_i^{(t-1)} = \mathcal{P}^{(t-1)}\left(A_{i}^{(t)}\right) \,.
\end{equation}
This iteration is initialized with $\varphi_i^{(T)}= 1$ and iterated down to layer $0$, where the resulting ACE is computed as
\begin{equation}
\varphi_i = \lambda \varphi_i^{(0)} \,.
\end{equation}
For a scalar quantity like the energy, non-linear functions of several expansions may be computed
\begin{equation}
E_i = \mathcal{F}(\varphi_i ) \,,
\end{equation}
and
\begin{equation}
E = \sum_{i=1}^N E_i \,,
\end{equation}
where the sum is taken over all $N$ particles in the system.

The force on atom $k$ is obtained as
\begin{equation}
\pmb{F}_k = - \nabla_k E
\end{equation}
In analogy to the computation of $E_i$ we start by iterating for computing $\nabla_k E_i$. We note that 
\begin{equation}
\nabla_j \phi_{ji}^{(t)} = - \nabla_i \phi_{ji}^{(t)}
\end{equation}
and for all atoms $j$ in $\mathcal{N}_i$
\begin{equation}
\pder{ \mathcal{P}^{(t-1)}(A_{i}^{(t)}) }{\tilde{\phi}_{ji}^{(t)} } = \pder{ \mathcal{P}^{(t-1)}(A_{i}^{(t)}) }{A_{i}^{(t)} } \,.
\end{equation}
The derivatives
\begin{equation}
\pder{ \mathcal{P}^{(t-1)}(A_{i}^{(t)}) }{A_{i}^{(t)} } \,,
\end{equation}
can be obtained efficiently by recursive evaluation \cite{Lysogorskiy2021PACE} and will therefore not be discussed here. We next collect prefactors
\begin{equation}
p_i^{(t-1)} = \pder{ \mathcal{P}^{(t-1)}(A_{i}^{(t)}) }{A_{i}^{(t)} }  W C \,,
\end{equation}
and define
\begin{equation}
\pmb{f}_{ki}^{(t)} = 
 \left(\nabla_k \phi_{ki}^{(t)} \right)  \varphi_k^{(t)} \,,
\end{equation}
for $k \neq i$ and
\begin{equation}
\pmb{f}_{ii}^{(t)} = - \sum_k^{\mathcal{N}_i}  \pmb{f}_{ki}^{(t)} \,. 
\end{equation}
This allows us to compute the required derivative recursively as
\begin{equation}
\nabla_k \varphi^{(t-1)}_{i} = p_i^{(t-1)} \left( \pmb{f}_{ki}^{(t)}  + \sum_j^{\mathcal{N}_i} \phi_{ji}^{(t)}  \nabla_k  \varphi^{(t)}_{j} \right) \,, \label{eq:gradace}
\end{equation}
and
\begin{equation}
\nabla_k E_i = \pder{\mathcal{F}(\varphi_i )}{\varphi_i} \lambda \nabla_k \varphi^{(0)}_{i} \,.
\end{equation}  
Eq.(\ref{eq:gradace}) extends the range of the interaction with every iteration by $\rc$, so that a naive implementation requires a neighborlist for atoms within $T\rc$ from the root atom $i$. We show next how to avoid using large neighborlists and poorly scaling summations.

To this end we write out the recursion explicitly,
\begin{align}
\nabla_k E_i &= \pder{\mathcal{F}(\varphi_i )}{\varphi_i} \lambda p_i^{(0)} \bigg( \pmb{f}_{ki}^{(1)}  \nonumber \\
                  &\phantom{++}+ \sum_{j_1}^{\mathcal{N}_i} \phi_{j_1 i}^{(1)}    p_{j_1}^{(1)} \bigg( \pmb{f}_{k j_1}^{(2)}  \nonumber \\
                  &\phantom{++++}+ \sum_{j_2} ^{\mathcal{N}_{j_1}} \phi_{j_2 j_1 }^{(2)}     p_{j_2}^{(2)} \bigg( \pmb{f}_{k j_2}^{(3)}  \nonumber \\
                  &\phantom{++++++}+ \sum_{j_3}^{\mathcal{N}_{j_2}} \phi_{j_3 j_2}^{(3)}  p_{j_3}^{(3)} \bigg( \pmb{f}_{k j_3}^{(4)}  \nonumber \\
                  &\phantom{++++++++}+ \sum_{j_4}^{\mathcal{N}_{j_3}} \phi_{j_4 j_3}^{(4)}   p_{j_4}^{(4)} \bigg( \pmb{f}_{k j_4}^{(5)} + \dots \bigg) \bigg) \bigg) \bigg) 
\bigg) \,
\end{align}  
The iteration terminates with $\nabla_k  \varphi^{(T)}_{j} = 0$. The complexity of the recursive evaluation stems from the iterative summation, which in a direct implementation leads to a scaling as $\mathcal{N}^T$. This poor scaling can be avoided altogether by collecting and iterating prefactors. We define
\begin{align}
h_i^{(0)} &= \pder{\mathcal{F}(\varphi_i )}{\varphi_i} \lambda p_i^{(0)} \,,\\
h_i^{(t)} &=  \sum_j^{\mathcal{N}_i} h_j^{(t-1)} \phi_{i j}^{(t)}  p_{i}^{(t)} \,,
\end{align}
for $t = 1, \dots, T-1$. The evaluation of the prefactors is linear with neighbors $\mathcal{N}$ and linear with layers $T$. The expression for the force is given as
\begin{equation}
\pmb{F}_k = - \sum_i^{\mathcal{N}_{k}}  \sum_{t=0}^{T-1} h_i^{(t)} \pmb{f}_{ki}^{(t+1)} \,,
\end{equation}
which requires another summation over neigbors $\mathcal{N}$ and the $T$ layers.

Therefore the poor scaling with $\mathcal{N}^T$ summations has been transformed to linear scaling with effort proportional to $T\mathcal{N}$ for energy and force evaluation. However the iterative construction means that  $h_i^{(t)}$ needs to be available on neighboring atoms, for example on 'ghost' atoms in LAMMPS domain decomposition.

\subsection{Algorithm}

For the algorithm we assume that bonds are stored as $i \to j$ for quantities $\phi_{ji}$ and that the index $n_{j \to i}$ of the inverse bond $j \to i$ can be looked up, i.e. $n_{j \to i} = \mathrm{INV}( n_{i \to j})$. If this is not available, the values of $\phi_{ji}^{(t)}$ and $f_{ji}^{(t)}$need to be computed and stored for the inverse bonds, too. We furthermore assume that tensors can be attached to atoms and communicated to ghost atoms. Finally, we use the recursive evaluation of \lACE as implemented in Ref.~\onlinecite{Lysogorskiy2021PACE} for $\varphi_i^{(t)}$ and $p_i^{(t)}$ on each layer $t$ and for every atom $i$.

This facilitates the following algorithm:
\begin{algorithm}[H]
\caption{\label{algo:Grad} Gradients of graph ACE}
    \begin{algorithmic}
    
        \State Init $\varphi_i^{(T)} =1$ for  $i = 1, \dots, N$ 
        \State Init $\nabla_k \varphi_i^{(T)} =0$ for $i = 1, \dots, N$ 
        \For{ layers $t = T, T-1, \dots, 1$}
            \For{atoms $i = 1, \dots, N$}
                \State  $A_i^{(t)} = 0$ 
                \For  {neighbors $j = 1, \dots, \mathcal{N}_i$}
                    \State Evaluate $\phi_{ji}^{(t)}$
                    \State Evaluate $\tilde{\phi}_{ji}^{(t)}$
                    \State Evaluate $A_{i}^{(t)} = A_{i}^{(t)} + \tilde{\phi}_{ji}^{(t)}$
                \EndFor 
                \State Evaluate $\varphi_i^{(t-1)}, p_i^{(t-1)}$\
            \EndFor 
        \EndFor 
        
        \State $E = 0$
        \For { atoms $i = 1, \dots, N$ }
            \State Evaluate $\varphi_i = \lambda \varphi_i^{(0)}$ 
            \State Evaluate $E_i = \mathcal{F}(\varphi_i ) $ 
            \State Evaluate $h_i^{(0)} = \pder{\mathcal{F}(\varphi_i )}{\varphi_i} \lambda p_i^{(0)}$
            \State Evaluate $E = E + E_i$ 
        \EndFor 
        \State Communicate $p_i^{(1)}$ to 'ghost' atoms
        \State Communicate $h_i^{(0)}$ to 'ghost' atoms
        \State Init $\pmb{F}_i = 0$ for $i = 1, \dots, N$
        \For {{layers} $t = 1, 1, \dots, T-1$}
             \For {atoms $i = 1, \dots, N$}
              \State $h_i^{(t)} = 0$ 
              \For { {neighbors} $j = 1, \dots, \mathcal{N}_i$}
                \State {Look up} $n_{j \to i} = \mathrm{INV}( n_{i \to j})$
                \State Evaluate $h_i^{(t)} = h_i^{(t)} + h_j^{(t-1)} \phi_{i j}^{(t)}  p_{i}^{(t)} $\\
                \State Evaluate $f_{ji}^{(t)}$ \\
              \EndFor 
              \For {{neighbors} $k = 1, \dots, \mathcal{N}_i$}
                \State {Look up} $n_{k \to i} = \mathrm{INV}( n_{i \to k})$
                \State Evaluate $\pmb{F}_k = \pmb{F}_k - h_i^{(t)} f_{ki}^{(t+1)}$
              \EndFor 
             \EndFor 
             \State {Communicate} $p_i^{(t+1)}$ {to 'ghost' atoms}
             \State {Communicate} $h_i^{(t)}$ {to 'ghost' atoms}
        \EndFor 
    \end{algorithmic}
\end{algorithm}

\subsection{Relevant variables with explicit indices}

The basis functions are represented as
\begin{equation}
\phi_{jinplm}^{(t)} = R_{nl}^{(t)}(r_{ji}) Y_{lm}(\hat{\pmb{r}}_{ji}) \,,
\end{equation}
with parity $p = (-1)^l$. Gradients of the basis functions are available, for example, in Refs.\onlinecite{Drautz19,Lysogorskiy2021PACE,Bochkarev2022PACEmaker}. On each layer $t$ from these effective radial functions are computed
\begin{align}
\tilde{\phi}^{(t)}_{ij n plm} &= \sum_{\substack{n_1 n_2 \\ l_1 m_2 l_2 m_2}} W^{(t)} _{n l  n_1 p_1 l_1  n_2 p_2 l_2} \nonumber \\ 
& C_{l_1 m_1 l_2 m_2}^{lm} \phi_{ij n_1 p_1 l_1 m_1 }^{(t)} \varphi^{(t)}_{jn_2 p_2l_2 m_2} \,, 
\end{align}
from which the atomic base is obtained
\begin{equation}
A^{(t)}_{i n plm} = \sum_{j} \tilde{\phi}^{(t)}_{ij n plm} \,. 
\end{equation}
On each layer then a local \lACE is carried out as Eq.(\ref{eq:effacerotalt}),
\begin{align}
& \varphi^{(t-1)}_{jnplm} =  c^{(t-1,0)}_{nplm }  +  \sum_{n_1l_1m_1} c^{(t-1,1)}_{nplm p_1 n_1 l_1 m_1}   A^{(t)}_{j n_1 p_1 l_1 m_1} \nonumber \\
&+ \sum_{p_1 n_1 l_1 m_1 p_2 n_2 l_2 m_2}  c^{(t-1,2)}_{nplm p_1 n_1 l_1 m_1 p_2 n_2 l_2 m_2}   \nonumber \\
&\times A^{(t)}_{j n_1 p_1 l_1 m_1}  A^{(t)}_{j n_2 p_2 l_2 m_2} + \dots \nonumber \\
&+  \sum_{n_1 p_1 l_1 m_1 \dots n_P p_P l_P m_P}  c^{(t-1,P)}_{nplm n_1 p_1 l_1 m_1 \dots n_P p_P l_P m_P}   \nonumber \\ 
&\times A^{(t)}_{j n_1 p_1 l_1 m_1}  A^{(t)}_{j n_2 p_2 l_2 m_2} \dots  A^{(t)}_{j n_P p_P l_P m_P}\,, 
\end{align}
and on the last layer the resulting \gACE is computed, Eq.(\ref{eq:recfin}),
\begin{equation}
\varphi_{in} = \varphi^{(0)}_{in100} \,. 
\end{equation}
and the energy Eq.(\ref{eq:EF}) computed as
\begin{equation}
E_i = \calF(\ace_{i1}, \ace_{i2}, \ace_{i3}, \dots ) \,.
\end{equation}
For the gradients  we collect prefactors as
\begin{equation}
p_{\substack{i n p l m \\n_1 p_1 l_1 m_1 \\n_2 p_2 l_2 m_2}}^{(t-1)} =  \sum_{n_3 p_3 l_3 m_3} \pder{ \varphi_{i n p l m}^{(t-1)}}{A_{i n_3 p_3 l_3 m_3 }^{(t)} }  W^{(t)} _{n_3 l_3  n_1 p_1 l_1  n_2 p_2 l_2} C_{l_1 m_1 l_2 m_2}^{l_3 m_3 } \,.
\end{equation}
The prefactors enter the weights of effective pairwise force gradients
\begin{equation}
h_{\substack{i n_1 p_1 l_1 m_1 \\n_2 p_2 l_2 m_2}}^{(0)} = \sum_n \pder{\mathcal{F}(\varphi_{in} )}{\varphi_{in}} p_{\substack{i n 100 \\n_1 p_1 l_1 m_1 \\n_2 p_2 l_2 m_2}}^{(0)} \,,
\end{equation}
and
\begin{equation}
h_{\substack{i n_3 p_3 l_3 m_3 \\n_4 p_4 l_4 m_4}}^{(t)} = \sum_{\substack{n_1 p_1 l_1 m_1 \\n_2 p_2 l_2 m_2}} \left( \sum_j^{\mathcal{N}_i}    h_{\substack{j n_1 p_1 l_1 m_1 \\n_2 p_2 l_2 m_2}}^{(t-1)} \phi_{i j n_1 l_1 m_1 }^{(t)} \right) p_{\substack{i n_2 p_2 l_2 m_2 \\n_3 p_3 l_3 m_3 \\n_4 p_4 l_4 m_4}}^{(t)} \,.
\end{equation}
The weights are combined with the effective force gradients for $k \neq i$
\begin{equation}
\pmb{f}_{ki n_1 p_1 l_1 m_1 n_2 p_2 l_2 m_2}^{(t)} =  (\nabla_k \phi_{ki n_1 p_1 l_1 m_1 }^{(t)} ) \varphi^{(t)}_{k n_2 p_2l_2 m_2} \,,
\end{equation}
for the computation of the forces
\begin{equation}
\pmb{F}_k = - \sum_i^{\mathcal{N}_{k}}   \sum_{\substack{n_1 p_1 l_1 m_1 \\n_2 p_2 l_2 m_2}} \sum_{t=0}^{T-1} h_{\substack{i n_1 p_1 l_1 m_1 \\n_2 p_2 l_2 m_2}}^{(t)} \pmb{f}_{ki n_1 p_1 l_1 m_1 n_2 p_2 l_2 m_2}^{(t+1)} \,.
\end{equation}

%

\end{document}